\documentclass{emulateapj}
\usepackage{graphicx}

\newcommand\refsec[1]{\S \ref{sec:#1}}
\newcommand\refeq[1]{eq.~(\ref{eq:#1})}
\newcommand\Rsrc{R_{\rm src}}
\newcommand\Rein{R_{\rm ein}}

\newcommand\mucut{\mu_{\rm cut}}

\newcommand{\be}{\begin{equation}}
\newcommand{\ee}{\end{equation}}
\newcommand{\ba}{\begin{eqnarray}}
\newcommand{\ea}{\end{eqnarray}}
\newcommand{\bcn}{\begin{center}}
\newcommand{\ecn}{\end{center}}
\newcommand{\ben}{\begin{enumerate}}
\newcommand{\een}{\end{enumerate}}
\newcommand{\btab}{\begin{tabular}}
\newcommand{\etab}{\end{tabular}}
\newcommand{\bt}{\begin{table}}
\newcommand{\et}{\end{table}}

\newcommand\reffig[1]{Figure \ref{fig:#1}}

\newcommand{\bfig}{\begin{figure}}
\newcommand{\efig}{\end{figure}}
\newcommand\bp{\begin{figure}}
\newcommand\ep{\end{figure}}
\newcommand\bpm{\begin{figure*}}
\newcommand\epm{\end{figure*}}

\newcommand {\apgt} {\ {\raise-.5ex\hbox{$\buildrel>\over\sim$}}\ }
\newcommand {\aplt} {\ {\raise-.5ex\hbox{$\buildrel<\over\sim$}}\ }
\newcommand\fib{{\rm fib}}

\newcommand\vx{{\bf x}}

\begin{document}

\title{Lensing Probabilities for Spectroscopically Selected Galaxy-Galaxy
Strong Lenses}

\author{
Gregory Dobler\altaffilmark{1,5},
Charles R. Keeton\altaffilmark{2},
Adam S. Bolton\altaffilmark{1,3},
and
Scott Burles\altaffilmark{4}
}

\altaffiltext{1}{
Harvard-Smithsonian Center for Astrophysics,
60 Garden Street, Cambridge, MA 02138 USA
}
\altaffiltext{2}{
Department of Physics and Astronomy, Rutgers University,
136 Frelinghuysen Road, Piscataway, NJ 08854 USA
}
\altaffiltext{3}{
B.W. Parrent Fellow,
Institute for Astronomy, University of Hawaii, 
2680 Wodlawn Dr., Honolulu, HI 96822, USA
}
\altaffiltext{4}{
Kavli Institute for Astrophysics and Space Research and
Department of Physics, Massachusetts Institute of Technology,
Cambridge, MA 02139 USA
}
\altaffiltext{5}{
gdobler@cfa.harvard.edu
}

\begin{abstract}
Spectroscopic galaxy-galaxy lens searches are presently the most prolific method 
of identifying strong lens systems in large data sets.  We study the 
probabilities associated with these lens searches, namely the probability of 
identifying a candidate with rogue [\ion{O}{2}] emission lines in a galaxy's
spectrum, and the probability that the candidate will show features of strong
lensing in follow-up photometric observations.  We include selection effects
unique to spectroscopic data, and apply them to the Sloan Lens ACS (SLACS)
survey \citep{SLACS1}.  The most significant selection effect is the 
finite size of the spectroscopic fiber which selects against large 
separation lenses and results in a non-monotonic dependence of the rogue line 
probability on velocity dispersion.  For example, with the 3 arcsec diameter 
SDSS fiber and 2 arcsec FWHM seeing, we find that the probability that a given LRG 
has a rogue [\ion{O}{2}] line in its spectrum \emph{decreases} with velocity 
dispersion from 150 km/s to 300 km/s and then increases up to 400 km/s for a given 
source size.  The total probability for observing a rogue line in a single survey 
spectrum is $\sim$0.9--3.0\%, and the total lensing rate is $\sim$0.5--1.3\%.  
The range is due to uncertainties in the physical size of [\ion{O}{2}] emission 
regions, and in the evolution of the [\ion{O}{2}] luminosity function.  Our 
estimates are a factor of $\sim$5 higher than the results of the SLACS survey, a 
discrepancy which we attribute to the SLACS requirement that multiple rogue lines 
be observed simultaneously.
\end{abstract}


\section{Introduction}
\label{sec:intro}

Spectroscopic gravitational lens searches have begun to yield a remarkable 
number of strong galaxy-galaxy (g-g) lens systems 
\citep{bolton04,SLACS1,willis}.  These finite source lenses promise both new 
physical insights and new phenomenology.  The extended images provide extensive 
constraints on the lens potential, especially on the radial density profile, 
which is still the 
main systematic uncertainty in lensing constraints on the Hubble constant
\citep[e.g.,][]{keko97,csk02}.  In present surveys, limits on the source
redshift range mean the lenses that are found typically have images that
appear well within the effective radius of the lens galaxy \citep{SLACS3}.
This makes g-g lenses ideal for probing the inner regions of distant
elliptical galaxies.  In addition, current selection effects favor
star-forming source galaxies, which opens the exciting possibility of
observing multiply-imaged supernovae \citep{og1,dk1}.  

The basic premise behind spectroscopic lens searches is to mine large
samples of galaxy spectra looking for ``rogue'' emission lines that
originate from background galaxies at small impact parameter 
\citep{warren96,willis00}.  This technique is complementary to photometric 
searches \citep[e.g.,][]{cabanac07,kubo07} which look for strongly lensed, 
arc-like features in imaging data.  

Among several recent spectroscopic searches \citep{bolton04,SLACS1,willis}, the 
most prolific has been the Sloan Lens ACS (SLACS) 
survey.\footnote{www.slacs.org}  For this survey, \citet{bolton04} 
and \citet{SLACS1} mined a catalog of 50,996 Sloan Digital
Sky Survey (SDSS) Luminous Red Galaxy \citep[LRG, see][]{LRG} spectra
for rogue [\ion{O}{2}] 3727 emission lines, and found $\sim$50 candidates.  The 
addition of later SDSS data releases as well as spectra from the MAIN galaxy 
sample increased the number of candidates to $\sim$200, with a similar 1-in-1000 
incidence.  Follow-up observations have subsequently confirmed 70--80 new g-g 
lenses from among these candidates.  SLACS data have been used to place the lens 
galaxies on the fundamental plane \citep{SLACS2,bolton-FP}, to constrain the 
redshift evolution of the density profiles of elliptical galaxies 
\citep{SLACS3}, and to trace the density profiles out to very large radii 
\citep{gavazzi07}.

Given that galaxy-galaxy lenses are already numerous, and will become
increasingly common in large surveys \citep{marshall,moustakas}, a sound 
statistical analysis of the expected incidence of g-g strong lensing is warranted.  
In traditional analyses of the statistics of lensed quasars \citep[e.g.,][]{tog}
the primary statistical question is, ``what is the probability
that a given source is lensed?''  By contrast, in g-g lens statistics
the question is different, viz.\ ``what is the probability that a given
galaxy is a lens?''  In this paper we formulate a general statistical
analysis applicable to spectroscopic g-g lens searches, and apply it
to the SLACS sample to estimate the total number of rogue emission
lines in the survey, and the actual number of strong lens systems that
should be confirmed by follow-up observations.  Our results for SLACS
will help assess the completeness of that survey, while our general
conclusions will (we hope) be useful in the design of future
spectroscopic lens searches.

Except where noted, throughout this paper we assume a flat cosmology with 
$\Omega_M = 0.3$, $\Omega_{\Lambda} = 0.7$, and $H_0 = 70$ km/s/Mpc.

\section{Probability for Galaxy-Galaxy Lensing}
\label{sec:theory}

\begin{figure*}
  \begin{center}
  \centerline{
  \includegraphics[width=0.9\textwidth]{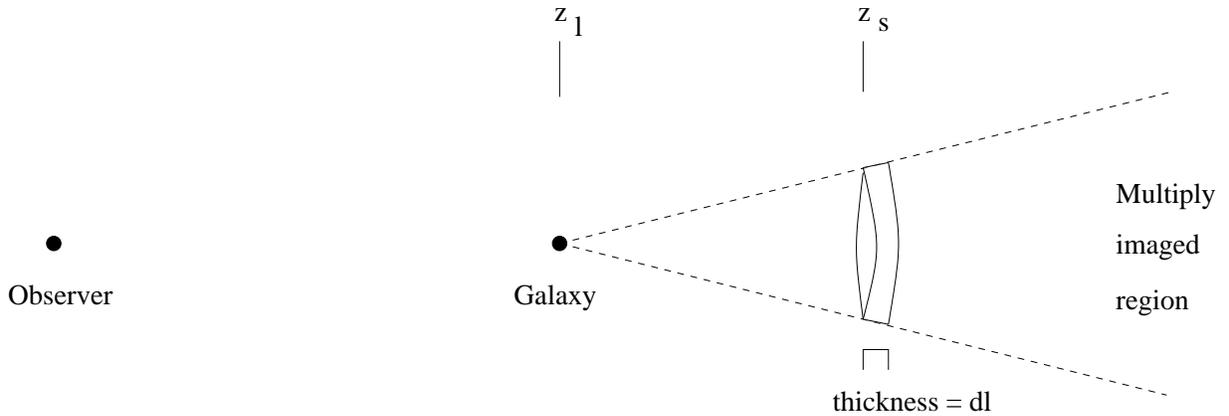}
  }
  \caption{
Lensing geometry for spectroscopically-selected galaxy-galaxy lenses.
The probability that the foreground galaxy is a lens is the probability
that there is a source within the galaxy's ``Einstein cone.''  (The
multiply-image region behind the galaxy is not strictly a cone, because
the cross section need not be circular and does not grow linearly with distance; 
but the terminology is attractive.)
}
  \label{fig:lensgeom}
  \end{center}
\end{figure*}

\reffig{lensgeom} shows a schematic representation of the lensing 
geometry.  Let the galaxy be described by a singular isothermal ellipsoidal (SIE) 
mass distribution with parameters $\vec{G} = (z_l, \sigma, e, \gamma, 
\phi_\gamma)$ where $z_l$ is the redshift, $\sigma$ is the velocity dispersion, 
$e$ is the ellipticity, $\gamma$ is the external shear, and $\phi_\gamma$ is
the angle between the ellipticity and shear.  The probability
$P_G(\vec{G})$ that this galaxy is a lens is equivalent to the
probability that there is a source within the ``Einstein cone''
of the galaxy.  Here we define the Einstein cone to be the region
behind the galaxy in which a source is strongly lensed.  For point
sources, the Einstein cone is the same as the multiply-imaged region,
but we will refine the definition shortly to incorporate complexities
from extended sources.  Note that the cross section of this region
need not be circular, and its size does not grow linearly with
distance, so the volume is not strictly conical; but we believe the
terminology is convenient and attractive.

If the number density of sources brighter than flux $S$
as a function of redshift is given by $n_s(z_s,S)$, then the lensing
probability is
\ba
  P_G(\vec{G}) &=& \int_{V_{\rm ein}} n_s(z_s,S)\ dV
    \label{eq:lensprob1}\\
  &=& \int_{z_l}^{\infty} \frac{dV}{dz_s d\Omega}\ dz_s 
    \int n_s(z_s,S)\ d\vec{u} ,
    \nonumber
\ea
where $V_{\rm ein}$ is the volume of the Einstein cone, $z_l$
and $z_s$ are the lens and source redshifts respectively, and $\Omega 
= \Omega(z_s)$ is the solid angle subtended by the cone at $z_s$.   
The integral over $\Omega$ is actually an integral over 
$\vec{u}$, the angular coordinates in the source plane, and for now
we consider the source plane integral to extend over the
multiply-imaged region.  Finally, it is natural to do the integral
in comoving coordinates, but to express distances as angular
diameter distances.  So we write the comoving volume element as
\be
  \frac{dV}{dz_s d\Omega} = \frac{c}{H_0}\ \frac{(1+z_s)^2 D_s^2}{E(z_s)}\ ,
\ee
where $E(z) = [\Omega_M (1+z)^3 + \Omega_{\Lambda}]^{1/2}$, and write
factors of $(1+z_s)$ explicitly allow us to keep $D_s$ as the
angular diameter distance to the source.

In practice, we are interested in the number density of sources
whose observed flux is above a survey's flux limit $S_0$.  This
implies $S_I \times \mu \geq S_0$ where $S_I$ is the source's
intrinsic flux and $\mu$ is the lensing magnification.  Therefore,
the relevant number density for \refeq{lensprob1} is
\be
  n_s(z_s,S) = n_s(z_s,S_0/\mu)
  = \int_{L_0/\mu}^{\infty} \Phi(L,z_s) \ dL ,
  \label{eq:ns}
\ee
where $\Phi(L,z_s)$ is the source luminosity function at
redshift $z_s$.  Here $L_0 = 4\pi (1+z_s)^4 D_s^2 S_0$ is the
luminosity corresponding to the flux limit $S_0$, and factors
of $(1+z_s)$ again appear so that we may keep $D_s$ as an
angular diameter distance.  The fact that the lower limit of
integration depends on $\mu$ means that the integral
automatically incorporates lensing magnification bias.

With point sources the definition of a lens relies on image
multiplicity: any source with multiple images is said to be
strongly lensed.  With extended sources the situation is more
complicated.  A source lying just outside the caustics might
be distorted enough to be labeled a lens even if there is just
one image.  A source lying astride a cusp or fold caustic may
exhibit a single arc comprising two or three merged images,
with counter-images that may or may not be bright enough to be
detectable.  The point is that identifying an object as a lens
may depend on some qualitative interpretation of the morphology.
Since the interpretation depends on distortions of the image(s),
which are related to the lensing magnification, we attempt to
quantify the labeling of extended lenses through a magnification
cut.  Specifically, we label an object a lens if the lensing
magnification $\mu$ exceeds some threshold set by $\mu_{cut}$.
(Our choice of $\mu_{cut}$ is discussed below.)  We then take
the source plane integral in \refeq{lensprob1} to extend over
the region in which $\mu > \mu_{cut}$.

The lensing magnification depends not only on the source
position $\vec{u}$, but also on the parameters of the lens
galaxy.  Therefore \refeq{lensprob1} represents the lensing
probability for a particular galaxy.  To describe a population
of galaxies, in principle we want to average over some appropriate
distributions of $z_l$, $\sigma$, $e$, $\gamma$, and $\phi_\gamma$.
In practice, if we have a set of $N$ observed galaxies, each of
which is described by the parameters $\vec{G}_i$, we can do the
average explicitly:
\be
  \bar{P}_G = \frac{1}{N} \sum_{i=1}^{N} P_G(\vec{G}_i) .
  \label{eq:totprob}
\ee
In the following sections we apply this $P_G$ calculation to the
SLACS sample, specifically incorporating parameters and selection
effects appropriate to that survey.

\section{SLACS Survey Parameters}
\label{sec:SLACS}

\subsection{Spectroscopic Selection Effects}

\begin{figure*}
  \begin{center}
  \centerline{
  \includegraphics[width=0.45\textwidth]{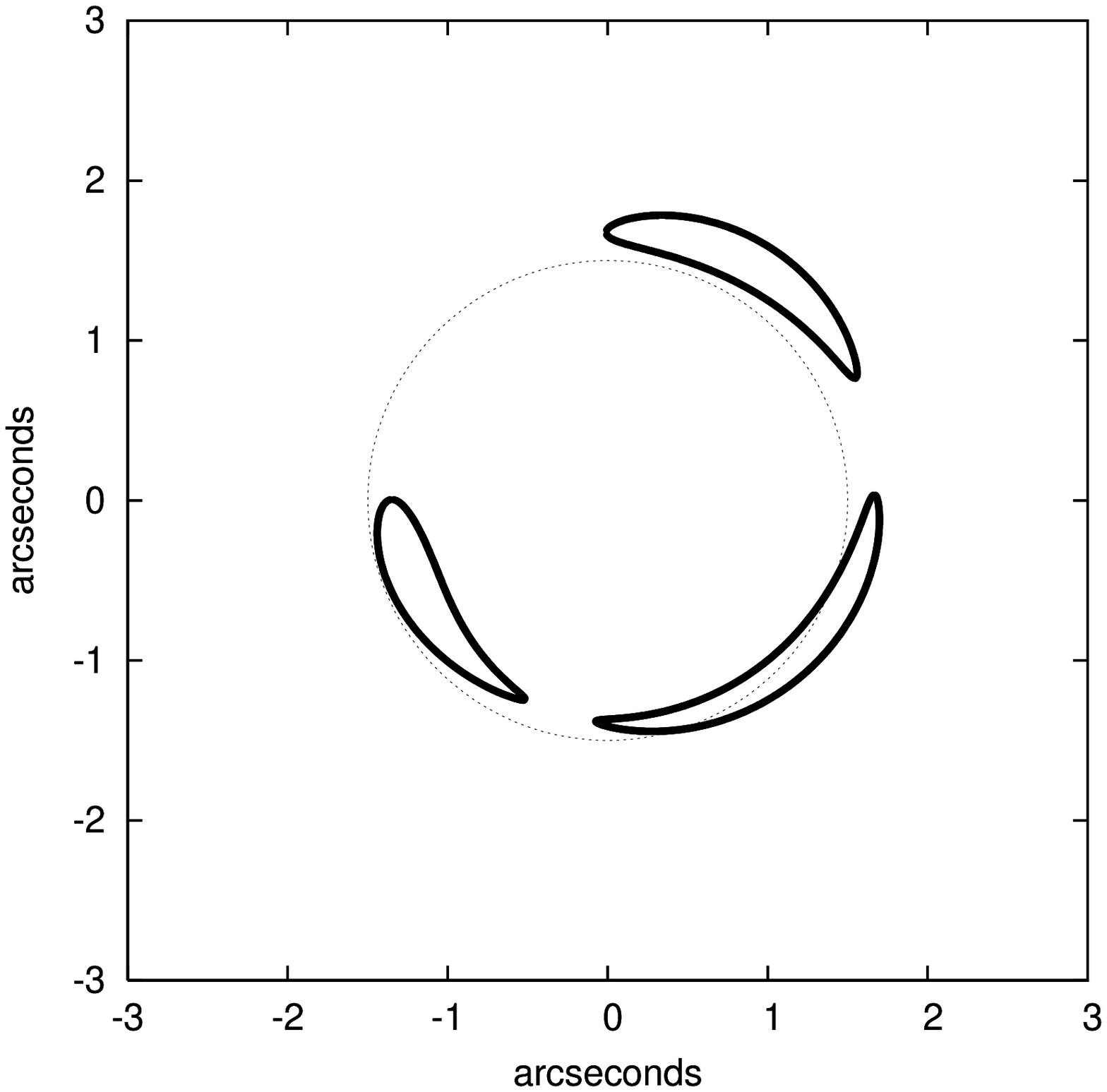}
  \includegraphics[width=0.45\textwidth]{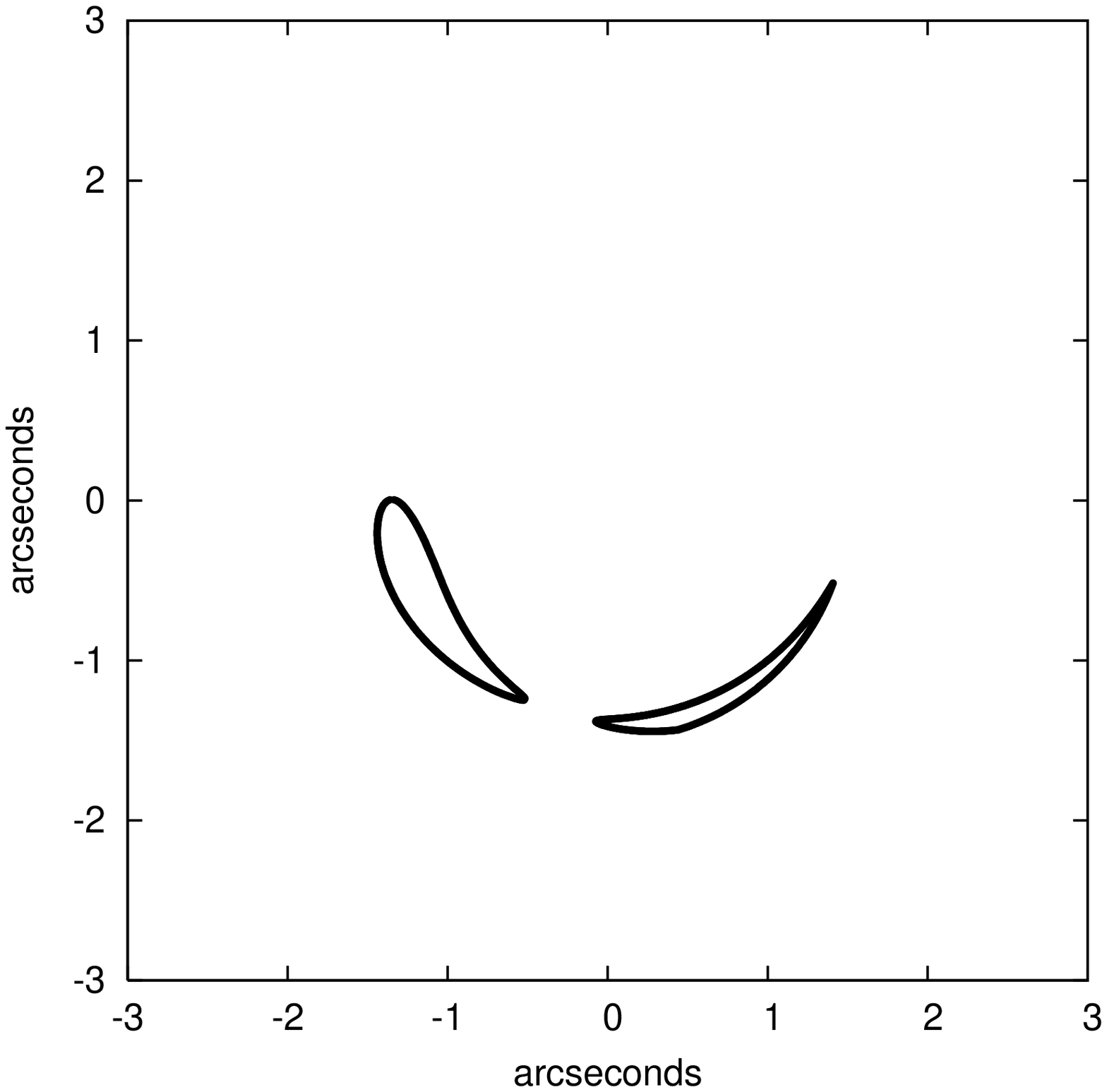}
  }
  \caption{
Selection effects due to the finite SDSS spectroscopic fiber size.
{\em (Left)}
The solid curves show a sample image configuration for a lens system
with lens parameters 
$(z_l,\sigma,e,\gamma,\phi_{\gamma}) = (0.2,300\mbox{ km/s},0.3,0.05,0.0)$
and source parameters $(z_s,\Rsrc,u_0,v_0)=(0.6,0.88\mbox{ 
kpc},0.15\arcsec,0.15\arcsec)$.
The dotted circle denotes the SDSS fiber.
{\em (Right)}
The solid curves show only the portion of the images that fall inside
the fiber.  The fiber cut reduces the integrated flux within the fiber,
creating a selection bias against large separation lenses and large
source sizes.  (This figure ignores the effects of seeing, which are
discussed in \refsec{seeing}.)
}
  \label{fig:fibcut}
  \end{center}
\end{figure*}

When discussing SDSS spectra, it is crucial to account for the finite
size of the spectroscopic fiber.  The strength of a rogue emission line
is directly related to how much flux from the source galaxy falls within
the fiber.  If a lens galaxy is massive, the lensed images may be pushed
outside of the fiber, so the rogue line may be weak or even absent
despite the fact that the system is a lens.  \reffig{fibcut} shows a
sample image configuration both with and without this ``fiber cut''
taken into account.  (This figure ignores the effects of seeing, which are 
discussed in \refsec{seeing}.)

The diameter of the SDSS fiber is 3.0 arcsec, while an SIE galaxy 
lens with redshift $z_l = 0.2$ and velocity dispersion
$\sigma = 250$ km/s (roughly the mean LRG redshift and velocity 
dispersion from the SLACS survey) has an Einstein radius of 
$\sim$1.0--1.2 arcsec depending on $z_s$.  Thus, the fiber cut may 
create a significant bias against large separation lenses in the SLACS 
sample.  There is also a 
bias against large sources (for a given source flux).  Of course, with 
follow up observations the full image configuration will be observed, 
but it is the fiber flux that determines whether a system is 
identified as a lens candidate in the first place.

The finite wavelength range of SDSS spectra places an upper limit
on detectable source redshifts.  The quality of spectral noise
modeling is another important factor, since imperfect sky subtraction
can leave emission line residuals that are not modeled by the
eigenspectra used to fit LRGs \citep[see Figure 1 of][]{bolton04}.
\citet{bolton04} were careful to account for imperfect sky
subtraction at long wavelengths, allowing them to probe deep into
the 7000--9000 \AA\ range.  A third factor is the SLACS selection
criteria: LRG spectra were required to exhibit not only blended
[\ion{O}{2}] 3727 lines, but also two longer-wavelength ``secondary''
features.  The longest-wavelength secondary feature was [\ion{O}{3}]
5007, while the shortest was H$\beta$ with rest wavelength 4863 \AA.
These lead to an upper limit on the source redshift of
$z_{s,{\rm max}} \sim 9200 {\rm \AA} / 5007 {\rm \AA} - 1 = 0.84$ and
$z_{s,{\rm max}} \sim 9200 {\rm \AA} / 4863 {\rm \AA} - 1 = 0.89$,
respectively.  In our calculations we therefore take
$z_{s,{\rm max}} = 0.9$.  Finally, to ensure a significant lensing
probability in their sample, \citet{bolton04} only searched 
for sources with velocities more than 5000 km/s behind the LRG,
corresponding to a lower limit on the source redshift of
$z_{s,{\rm min}} = z_l + 0.017$.  These constitute the limits of
the $z_s$ integration in \refeq{lensprob1}.

The last important spectral parameter is the flux limit.  Figure~1
of \citet{bolton04} shows the typical 1$\sigma$ noise spectrum.
For wavelengths $\la$7200 \AA\ the wavelength dependence is small,
so for simplicity we take the noise floor to be constant in
wavelength, which also means constant in source redshift.
Including the \cite{bolton04} requirement that secondary emission
features have signal-to-noise ratios greater than 3.0, we take
$S_0 = 6.0 \times 10^{-17}$ ergs/s/cm$^2$.

\subsection{Source Population: \ion{O}{2} Luminosity Function}

From \refeq{ns} it is clear that we must specify the luminosity
function (LF) of sources, in order to account for the flux limit
and the magnification bias.  For SLACS, the primary selection is
on [\ion{O}{2}] line flux.  \citet{hogg} give the [\ion{O}{2}] LF
for the redshift range $0.3 < z < 1.5$ (see their Fig.~7), which
covers the range of source redshifts accessible in the SLACS
survey.  However, we make two refinements to the LF.  First,
\citeauthor{hogg} quoted the LF for an OCDM cosmology, but we
prefer to work in $\Lambda$CDM; luminosities and volumes 
both need to be adjusted.  Second, since [\ion{O}{2}] emission is 
thought to trace the star formation rate, the LF may vary 
substantially with redshift \citep{kenn, glaze}.  We include the 
possibility of number evolution by modeling the LF as an evolving 
Schechter function,
\be
  \Phi(L,z_s) \ dL = n_* (1+z_s)^{\beta} 
    \left(\frac{L}{L_*}\right)^{\alpha}\ e^{-L/L_*}\ \frac{dL}{L_*}\ .
  \label{eq:LF}
\ee

We make both adjustments using the following technique.  We first
choose a value for the evolution parameter $\beta$.  We postulate a
set of Schechter function parameters $(n_*,L_*,\alpha)$ to specify
the $\Lambda$CDM LF, $\Phi_{\Lambda{\rm CDM}}$.  We draw from this LF to generate
a mock sample of sources in an $\Lambda$CDM universe.  We then imagine
``observing'' these sources, interpreting them using an OCDM
cosmology, and deriving the effective OCDM LF, $\Phi'_{\rm OCDM}$.
We compare $\Phi'_{\rm OCDM}$ with the OCDM LF presented by
\citet{hogg} to see how well they match.  We then repeat this
process for many values of $(n_*,L_*,\alpha)$ and choose the
values that provide the best match between $\Phi'_{\rm OCDM}$ and
the \citeauthor{hogg} data.  This gives us the best-fit $\Lambda$CDM LF,
for our particular choice of the evolution parameter $\beta$.
Finally, we repeat the entire analysis for different values of
$\beta$.  The resulting LFs are summarized in Table \ref{tbl:LFfit}.

\begin{deluxetable}{cccc}
\tablewidth{0pt}
\tablecaption{Source \ion{O}{2} LF Parameters}
\tablehead{
  $n_*$ & $\log L_*$ & $\alpha$ & $\beta$ \\
  ($10^{-3}$ Mpc$^{-3}$) & (erg/s) & &
}
\startdata
   4.09 & 42.34 & \ -1.15 \ & 0 \\
   2.16 & 42.31 & \ -1.13 \ & 1 \\
   0.90 & 42.37 & \ -1.17 \ & 2 \\
   0.62 & 42.18 & \ -1.09 \ & 3
\enddata
\tablecomments{
Best-fit Schechter function parameters for the \ion{O}{2} lumonisity
function in the redshift range $0.3<z<1.5$.  We have converted the
data of \citet{hogg} from OCDM to $\Lambda$CDM, and we have considered
different possibilities for the number evolution parameter $\beta$
(see text).
}\label{tbl:LFfit}
\end{deluxetable}

With an LF of the form \refeq{LF}, the luminosity integral in
\refeq{ns} can be evaluated,
\be
  \int_{L_0/\mu}^{\infty} \Phi(L,z_s) \ dL
  = n_*\, (1+z_s)^{\beta}\, \Gamma\left[1+\alpha,\frac{L_0}{L_*\mu}\right] ,
  \label{eq:numsrc}
\ee
where $\Gamma$ is the incomplete gamma function.  Recall that for
a flux-limited survey, $L_0 = 4\pi (1+z_s)^4 D_s^2 S_0$ depends on
redshift.

\subsection{Lens Population: SDSS LRG Sample}

The initial sample analyzed by \citet{bolton04} included $\sim$51,000
LRG spectra obtained by SDSS between 2000 March 5 and 2003 May 27.  To
obtain proper statistics, we must include appropriate distributions
of $z_l$, $\sigma$, $e$, $\gamma$, and $\phi_{\gamma}$ for this
sample (see eq.~\ref{eq:totprob}).  The velocity dispersion function
$dn/d\sigma$ of the full SDSS elliptical galaxy catalog has been
measured by \citet{sheth03}.  Their analysis corrected for various
selection effects in order to recover the intrinsic distribution
$dn/d\sigma$.  However, we wish to \emph{include} the selection
effects since our goal is to estimate how many rogue emission lines
should have been found in the actual SDSS data.

To do this, and also to account for distributions of $z_l$ and $e$,
we randomly select 800 LRGs \citep[see][]{LRG} observed between
2000 March 5 and 2003 May 27 and flagged as GALAXY-RED by the SDSS
photometric pipeline \citep{lupton}.\footnote{For details related to
SDSS, see \citet{SDSS1} for a technical summary, \citet{SDSS2} for
issues related to the camera, \citet{SDSS3}, \citet{SDSS4}, and
\citet{SDSS5} for a discussion of the photometric system and
calibration, and \citet{SDSS6} for details related to astrometric
calibration.  The tiling procedure is described in \citet{SDSS7}.}
The number of LRGs was chosen to be computationally tractable, and we have 
verified that it is a sufficiently large sample to yield accurate statistics (see 
\refsec{totprob}).
Choosing from the sample of observed LRGs automatically
incorporates all of the same selection effects as the sample from
which SLACS was drawn.  As in \cite{bolton04}, we restrict the LRG
redshift range to $0.15<z_l<0.65$.\footnote{The distributions of galaxies in
redshift and velocity dispersion are shown in \reffig{histograms}.}
We use $r$-band ellipticities
from de Vaucouleurs fits in the SDSS photometric pipeline
\citep{lupton}.  These ellipticities describe the light while
what we really need is the mass, but there is evidence that the
mass and light ellipticities follow similar distributions
\citep{rusinteg,heyl,nabbur}.

We assign each galaxy a random shear amplitude drawn from a
lognormal distribution centered on $\gamma = 0.05$ with dispersion
0.2 dex \citep[see][]{holdschech}, and a random shear angle
$\phi_{\gamma} \in [0,2\pi]$.

\section{Methods}
\label{sec:methods}

Our formula for the total lensing probability explicitly includes
two integrals over the source redshift $z_s$ and the source position
$\vec{u}$ (see eq.~\ref{eq:lensprob1}).  There is a third integral
that enters implicitly: an integral over the image plane to compute
the magnification of an extended source.  The integral over source
redshift is straightforward to compute numerically, but the 2-D
integrals over the image and source planes require more care.

\subsection{Semi-Analytic Image Plane Integration}
\label{sec:sieanalytic}

To calculate the magnification $\mu$ for a given source and lens,
we extend the analytic method developed in \citet{dobler} to include
finite source lensing by isothermal \emph{ellipsoids} (SIEs) in an
external shear field.  The SIE density profile has been used quite
successfully to model not only quasar lenses but also the extended
images seen in SLACS lenses \citep{SLACS3}.  The lens equation is
\be
  \vec{u} = \left( \begin{array}{cc}
    1 - \gamma\cos 2\phi_{\gamma} & -\gamma\sin 2\phi_{\gamma} \\
    -\gamma\sin 2\phi_{\gamma} & 1 + \gamma\cos 2\phi_{\gamma}
  \end{array} \right) \vec{x} - \vec{\alpha}(\vec{x}) ,
\ee
where $\vec{x} = (r\cos\theta, r\sin\theta)$ are image plane
coordinates.  The two components of the deflection angle for an
SIE lens are \citep{kormann94,keeton98}
\ba
  \alpha_x &=& \frac{b' q}{\sqrt{1-q^2}}\ 
    \tan^{-1}\left[Q(\theta) \cos\theta\right] , \\
  \alpha_y &=& \frac{b' q}{\sqrt{1-q^2}}\ 
    \tanh^{-1}\left[Q(\theta) \sin\theta\right] ,
\ea
where $q=1-e$,
\ba
  Q(\theta) &=& \left(\frac{1-q^2}{q^2\cos^2\theta+\sin^2\theta}\right)^{1/2} ,
    \\
  b' &=& \frac{b\pi}{2 K(1+q^{-2})}\ ,
\ea
and $K$ is the complete elliptic integral of the first kind 
\citep[see][]{huterer}. The Einstein radius $b$ of the galaxy is related to its 
velocity dispersion by
\be
  b = 4\pi \left(\frac{\sigma}{c}\right)^2 \frac{D_{ls}}{D_s}\ ,
  \label{eq:rein}
\ee
where $D_s$ and $D_{ls}$ are angular diameter distances from the
observer to the source and from the lens to the source, respectively.

The idea behind our analytic method is to parameterize the source
boundary by a circle:
$(u,v) = (u_0 + \Rsrc\cos\lambda,v_0 + \Rsrc\sin\lambda)$ for
$\lambda \in [0,2\pi]$.  Plugging this into the lens equation yields
\ba
  u_0 + \Rsrc\cos\lambda &=& r\,\Gamma_- - \alpha_x, \nonumber\\
  v_0 + \Rsrc\sin\lambda &=& r\,\Gamma_+ - \alpha_y,
  \label{eq:lenseqSLACS}
\ea
with
\be
  \Gamma_{\pm} \equiv (1\pm\gamma\cos2\phi_{\gamma})\cos\theta 
  - \gamma\sin2\phi_{\gamma}\sin\theta.
\ee
We square and add the two equations in \refeq{lenseqSLACS} to
eliminate $\lambda$.  Since $\alpha_x$, $\alpha_y$, and $\Gamma_{\pm}$
are all independent of $r$, we obtain a simple quadratic equation for
$r$ that we can solve to find the following \emph{analytic} expression
for the image boundary as a function of $\theta$:
\be\label{eq:radii}
  r_{\pm}(\theta) = \frac{B \pm \sqrt{B^2 - A C}}{A}
  \label{eq:analyticsol}
\ee
where
\ba
  A &=& \Gamma_+^2 + \Gamma_-^2 , \nonumber\\
  B &=& \Gamma_-(\alpha_x+u_0) + \Gamma_+(\alpha_y+v_0) , \nonumber\\
  C &=& (\alpha_x+u_0)^2 + (\alpha_y+v_0)^2 - \Rsrc^2 .
\ea

The total magnification of an extended source is then
\be
  \mu_{\rm tot} = \frac{\mbox{Total image area}}{\mbox{Total source area}}
  = \frac{1}{2\pi \Rsrc^2} \int_{\mathcal{M}} \left( r_+^2 - r_-^2 \right)\ 
    d\theta
  \label{eq:muanalytic}
\ee
where $\mathcal{M}$ is the region of $\theta$ over which $r_{\pm}(\theta)$
is real and positive.  We can also impose the SDSS fiber cut very
simply as follows.  Let $r_{\rm fib} = 1.5$ arcsec be the fiber radius.
If we define the ``fiber magnification'' to be the total flux within
the fiber divided by the total flux of the source, we can compute
this as
\be\label{eq:noseemag}
  \mu_{\rm fib} = \frac{1}{2\pi \Rsrc^2} \int_{\mathcal{M}} 
    \left[ \min(r_+,r_{\rm fib})^2 - \min(r_-,r_{\rm fib})^2 \right]\ 
    d\theta
  \label{eq:mufib}
\ee
We emphasize that in both cases the solution for the image boundary
is completely analytic, which allows us to reduce the 2-D image
plane integration to a 1-D integral along the image boundary.

It is useful to note explicitly where the various system parameters
enter into the magnification calculation.  The lens and source redshifts
and the galaxy velocity dispersion all enter through the Einstein
radius $b$.  The shear parameters appear in $\Gamma_{\pm}$, while the
ellipticity is buried in the deflection components $\alpha_x$ and
$\alpha_y$.

\subsection{Seeing Effects}
\label{sec:seeing}

\begin{figure}
  \begin{center}
  \includegraphics[width=0.23\textwidth]{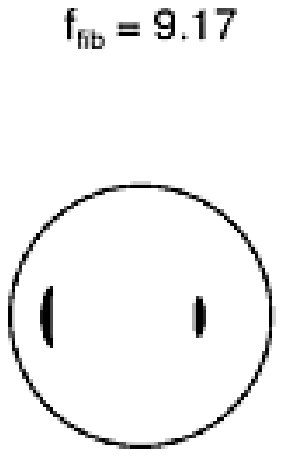}
  \includegraphics[width=0.23\textwidth]{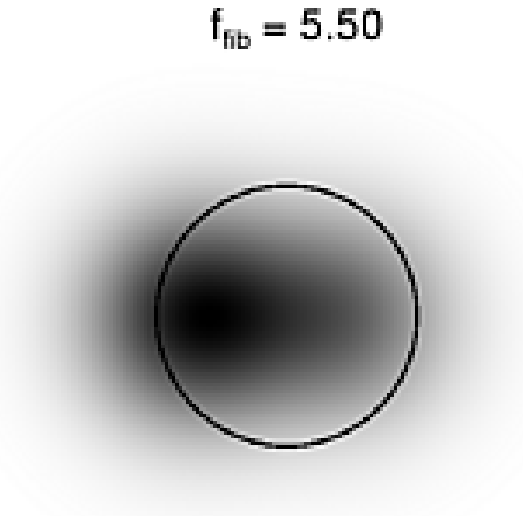}

  \includegraphics[width=0.23\textwidth]{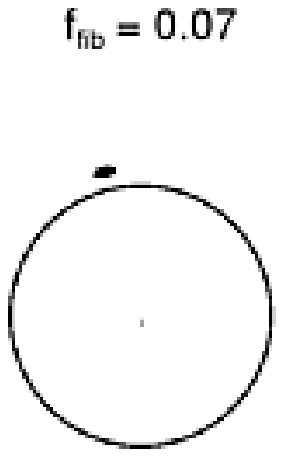}
  \includegraphics[width=0.23\textwidth]{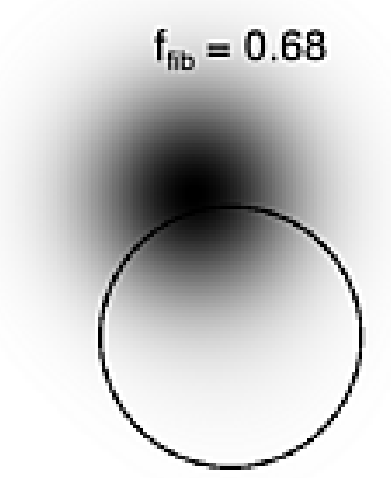}
  \caption{
Sample image configurations for on-axis (top) and off-axis (bottom) lensing 
geometries both with (right) and without (left) seeing.  Here $\Rein = 0.9$ 
arcsec and $\Rsrc=0.07$ arcsec, the source flux is normalized to unity, the
circle represents the $r_{\rm fib} = 1.5$ arcsec SDSS fiber, and the seeing
is 2 arcsec (FWHM).  For on-axis sourcees, the fiber flux is preferentially
\emph{decreased} due to seeing, while for off-axis sources the fiber flux
is preferentially \emph{increased}.
}
  \label{fig:seeing}
  \end{center}
\end{figure}

The above treatment of the image plane integration neglects the effects of 
atmospheric seeing.  For the SLACS sample, the seeing is typically 
$\sim$2 arcsec (FWHM), which is an appreciable fraction of the SDSS fiber
diameter.  Seeing can either \emph{add} flux to the fiber from images
outside, or \emph{remove} flux from the fiber from images that are
inside.  Which of these effects dominates depends on the configuration
of images, as shown in \reffig{seeing}.

We handle the effects of seeing in the following way.  Let $I(\vx)$ be the surface 
brightness distribution of the lensed image (in the absence of seeing).  Then the 
smeared surface brightness distribution is
\be
  I'(\vx) = \int G(\vx-\vx')\ I(\vx')\ d\vx'
\ee
where $G$ represents the PSF, which we take to be a Gaussian with FWHM 2 arcsec.  
Specifically, $G(\vx-\vx')$ is the flux at $\vx$ when the PSF is centered at 
$\vx'$.  The fiber flux is then
\begin{eqnarray}
  F_\fib &=& \int_\fib I'(\vx)\ d\vx \nonumber\\
  &=& \int_\fib d\vx \int d\vx'\ G(\vx-\vx')\ I(\vx') \nonumber\\
  &=& \int d\vx'\ I(\vx') \left[ \int_\fib d\vx\ G(\vx-\vx') \right].
\label{eq:ffib}
\end{eqnarray}
The term in square brackets is the fiber flux of a Gaussian centered
at $x'$.  Since the Gaussian PSF and the fiber are both circular, this
term can only depend on the distance of the center of the Gaussian
from the origin, $r = |\vx'|$.  Hence we write this factor as $G_\fib(r)$,
and then rewrite \refeq{ffib} as
\be
  F_\fib = \int_{0}^{\infty} dr\,r \int_{0}^{2\pi} d\theta\ 
I(r,\theta)\ G_\fib(r).
\ee
If we take the source flux to be unity, the integral actually gives
the fiber magnification $\mu_\fib$.  In this case, the source surface
brightness is $1/\pi\Rsrc^2$, and since lensing conserves surface
brightness we have $I(x) = 1/\pi\Rsrc^2$ within the image boundaries,
and 0 outside.  Since the image boundaries are given by $r_\pm(\theta)$ from 
\refeq{radii}, in the end we can write the fiber magnification in the presence of 
seeing as
\be
  \mu_\fib = \frac{1}{\pi\Rsrc^2} \int d\theta \int_{r_-(\theta)}^{r_+(\theta)} 
dr\ r\ G_\fib(r),
\ee
which is a generalized version of \refeq{noseemag}.
The ``fiber Gaussian'' factor $G_\fib(r)$ must be computed numerically, however 
this 1D integral only needs to be done once making this semi-analytic method for 
the image plane integration computationally orders of magnitudes faster than 
classical ray shooting methods for computing the magnification of an extended 
source.

\subsection{Numerical Source Plane Integration}
\label{sec:srcint}

Since the magnification must be computed numerically, the source plane
integral in \refeq{lensprob1} must be computed numerically as well.
The numerical integration scheme we use incorporates multiple grids
with adaptive resolution to tile the source plane efficiently;
details are given in the Appendix.

\section{Results}
\label{sec:results}

\subsection{Three Probabilities}
\label{sec:three-prob}

As discussed in \refsec{theory}, we use a magnification cut to
determine whether a system is labeled a strong lens or not.  Since
the labeling is applied only after follow-up observations, we apply
the cut to the total magnification (not the fiber magnification).
We choose as our fiducial threshold $\mu_{\rm cut} = 2$, because
this corresponds to the magnification of a point source located on
the boundary of the multiply-imaged region for an isothermal sphere
lens.  A finite source magnified by this amount should show clear
signs of strong lensing.  In other words, we define
\be \label{eq:PL}
  P_L(\vec{G};\Rsrc) = P_G(\vec{G} \ | \ \mu_{tot}(\Rsrc)>\mu_{cut}) .
\ee
to be the probability that a galaxy has a rogue emission line due
to a source that is lensed.

We should not just discard systems with magnifications below the
cut.  There is a range of positions (roughly speaking, behind the
galaxy but outside the Einstein cone) where a source could send
enough flux down the fiber to create a rogue line without being
lensed.  Such systems represent false positives in spectroscopic
lens searches, and in order to understand the efficiency of a
survey we need to assess the false positive rate.  We define the
complement of \refeq{PL} to be the probability that a galaxy has
a rogue emission line due to a source that is {\em not} lensed:
\be \label{eq:PN}
  P_N(\vec{G};\Rsrc) = P_G(\vec{G} \ | \ \mu_{tot}(\Rsrc)<\mu_{cut}) .
\ee
We can then let $P_R = P_L + P_N$ be the total probability that
a galaxy has a rogue line, while $R_F = P_N/P_R$ is the false
positive rate (defined to be the fraction of candidates for which
follow-up observations do not show substantial evidence for lensing).
Throughout our analysis we keep track of the total rogue line
probability $P_R$, the lensing probability $P_L$, and the false
positive rate $R_F$.

\subsection{Back of the Envelope Estimate}

Before giving our full results, we can make a simple estimate of
the lensing probability.  For this estimate we ignore finite source
and fiber effects and just take all sources inside the Einstein cone
to be lensed.  For simplicity we take the luminosity threshold $L_0$
to be independent of source redshift, and we consider a non-evolving
source luminosity function, so $n_s$ is constant.  If we let $L_0$
be the lowest luminosity threshold (i.e., computed from the flux
limit $S_0$ at the minimum source redshift), we should overestimate
the number of detectable sources.  With these simplifications, the
lensing probability is
\ba
  P_L^{\rm est} &\sim& \int_{D^C_l}^{D^C_{s,{\rm max}}}  n_s\ \pi (D^C_s b)^2\ dD^C_s
    \nonumber\\
  &\sim& \int_{D^C_l}^{D^C_{s,{\rm max}}} n_s\ 16\pi^3 
      \left(\frac{\sigma}{c}\right)^4 (D^C_s-D^C_l)^2\ dD^C_s ,
    \nonumber\\
  &\sim& n_s\ \frac{16\pi^3}{3} \left(\frac{\sigma}{c}\right)^4 
    (D^C_{s,{\rm max}} - D^C_l)^3 .
    \nonumber\\
  &\sim& n_s\ \frac{16\pi^3}{3} \left(\frac{\sigma}{c}\right)^4 \times
    \nonumber\\
  & & \ \ \left[(1+z_{s,{\rm max}}) D_{s,{\rm max}} - (1+z_l) D_l\right]^3 ,
\ea
where $C$ indicates a comoving distance.  We take $z_{s,{\rm max}} = 0.9$.
For fiducial LRG parameters $z_l = 0.2$ and $\sigma = 250$ km/s, our
estimate yields $P_L^{\rm est} \sim 3.1\%$.  In other words, we expect
the lensing probability to be at the percent level.

\subsection{Dependence on Lens Parameters}

\begin{figure*}
  \begin{center}
  \centerline{
  \includegraphics[width=0.32\textwidth]{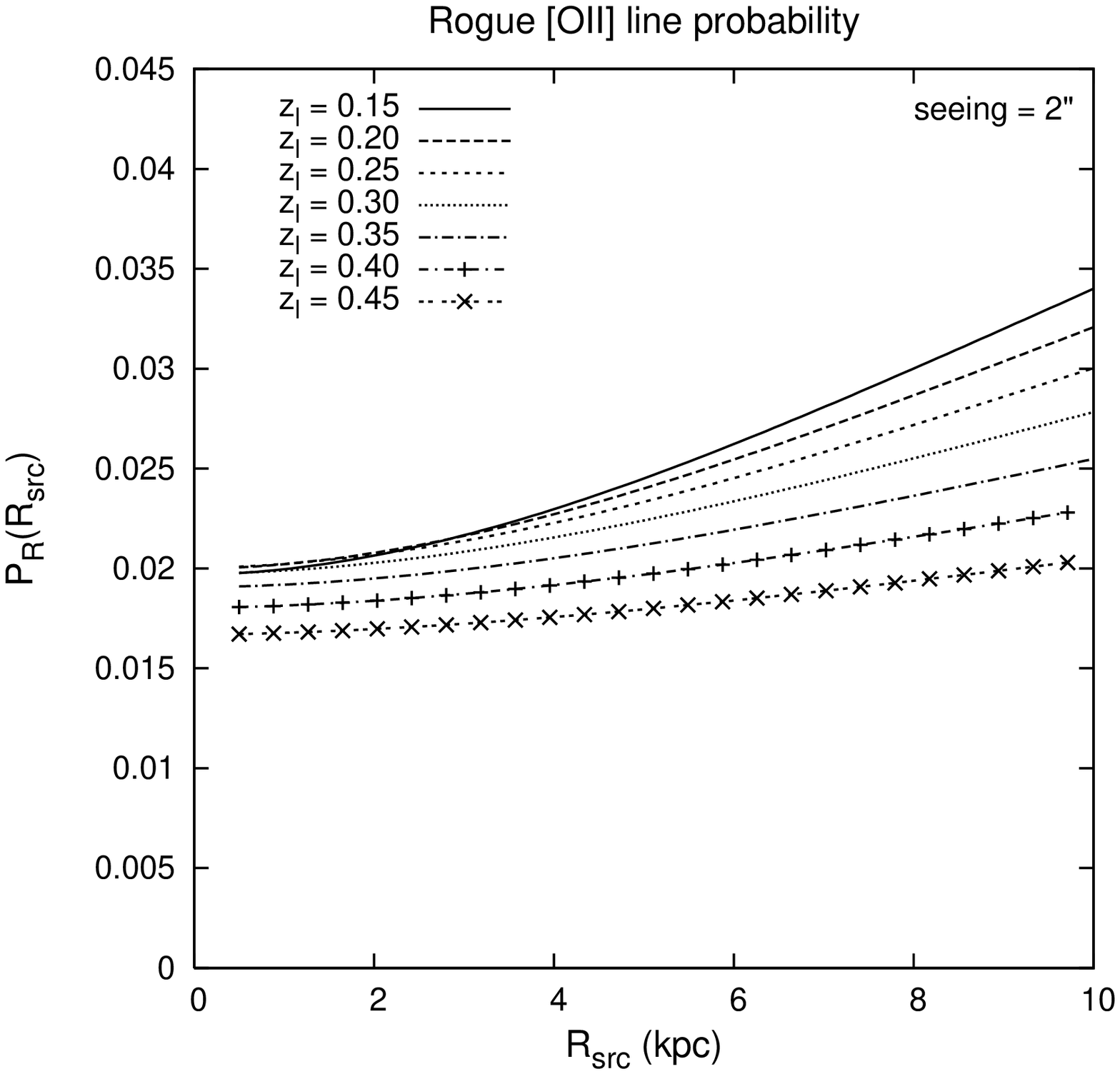}
  \includegraphics[width=0.32\textwidth]{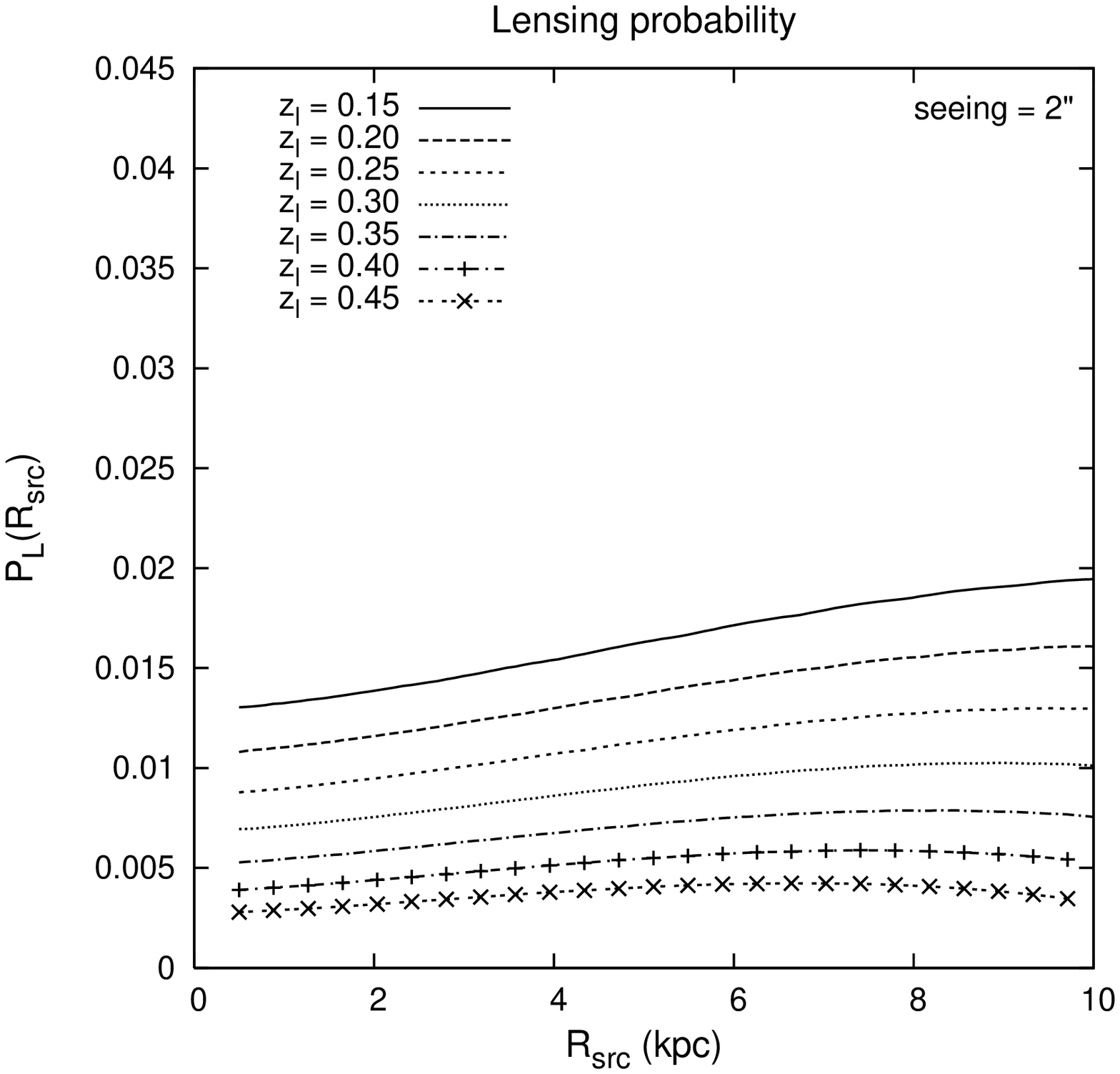}
  \includegraphics[width=0.32\textwidth]{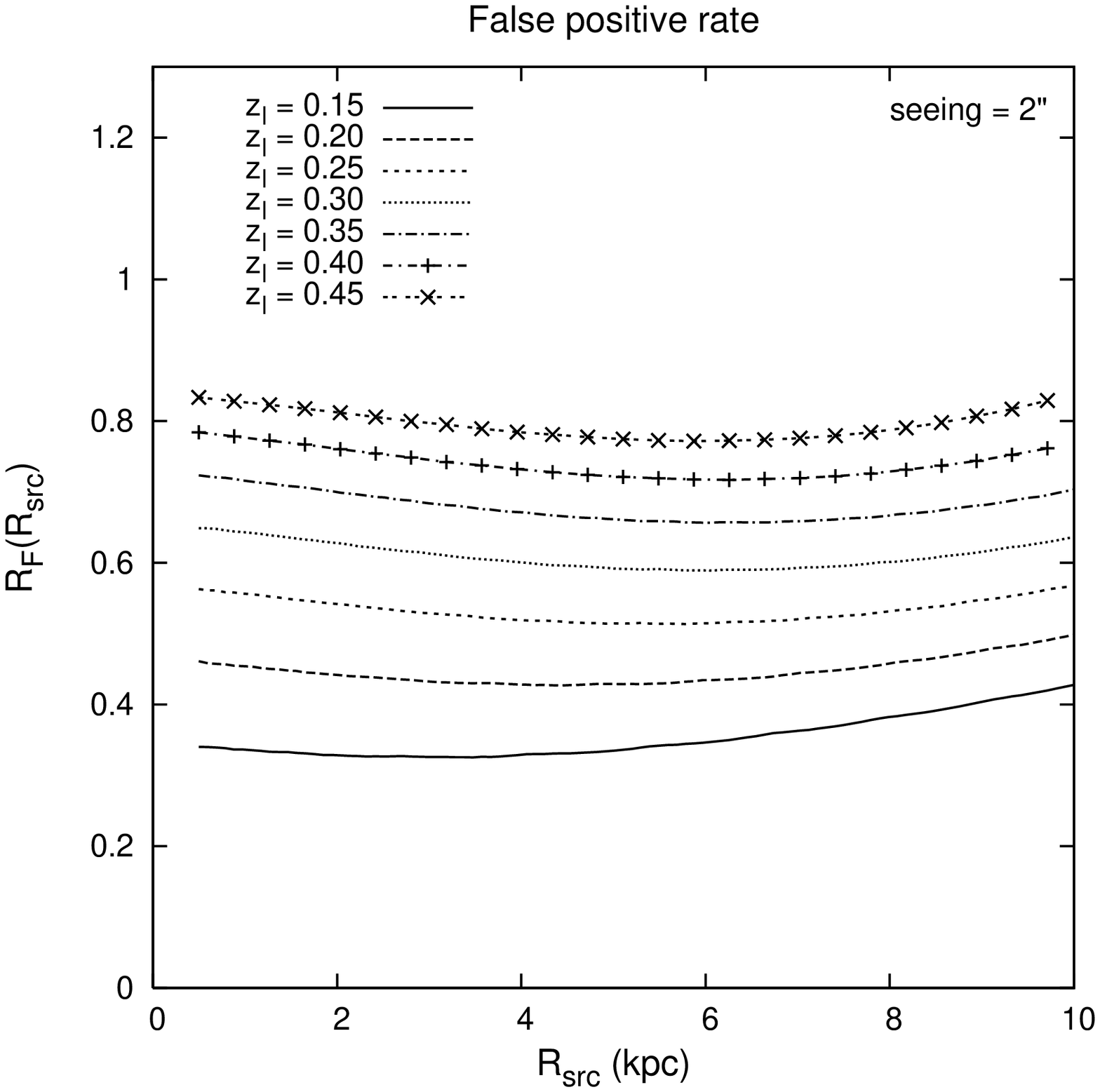}
  }
  \centerline{
  \includegraphics[width=0.32\textwidth]{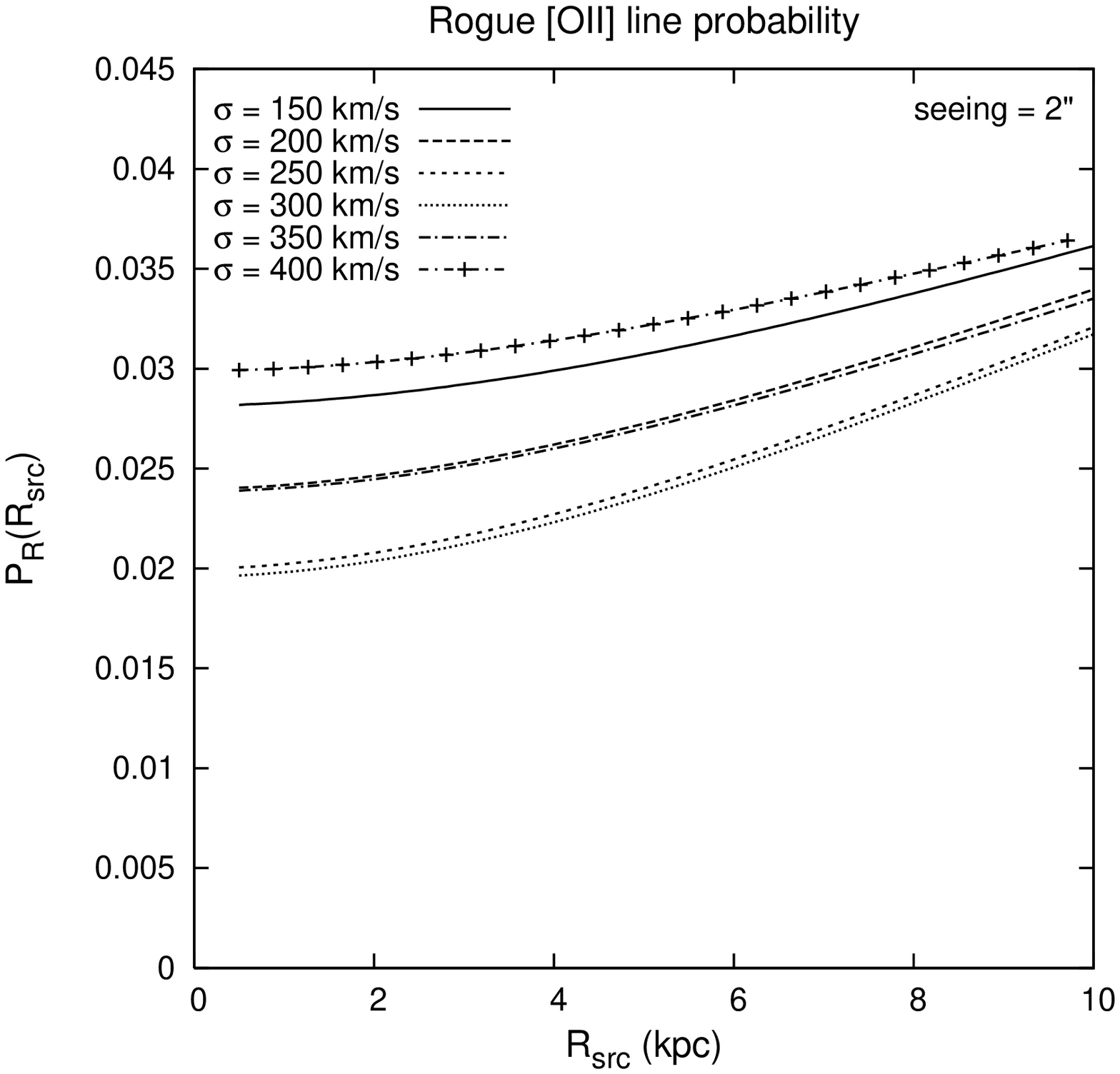}
  \includegraphics[width=0.32\textwidth]{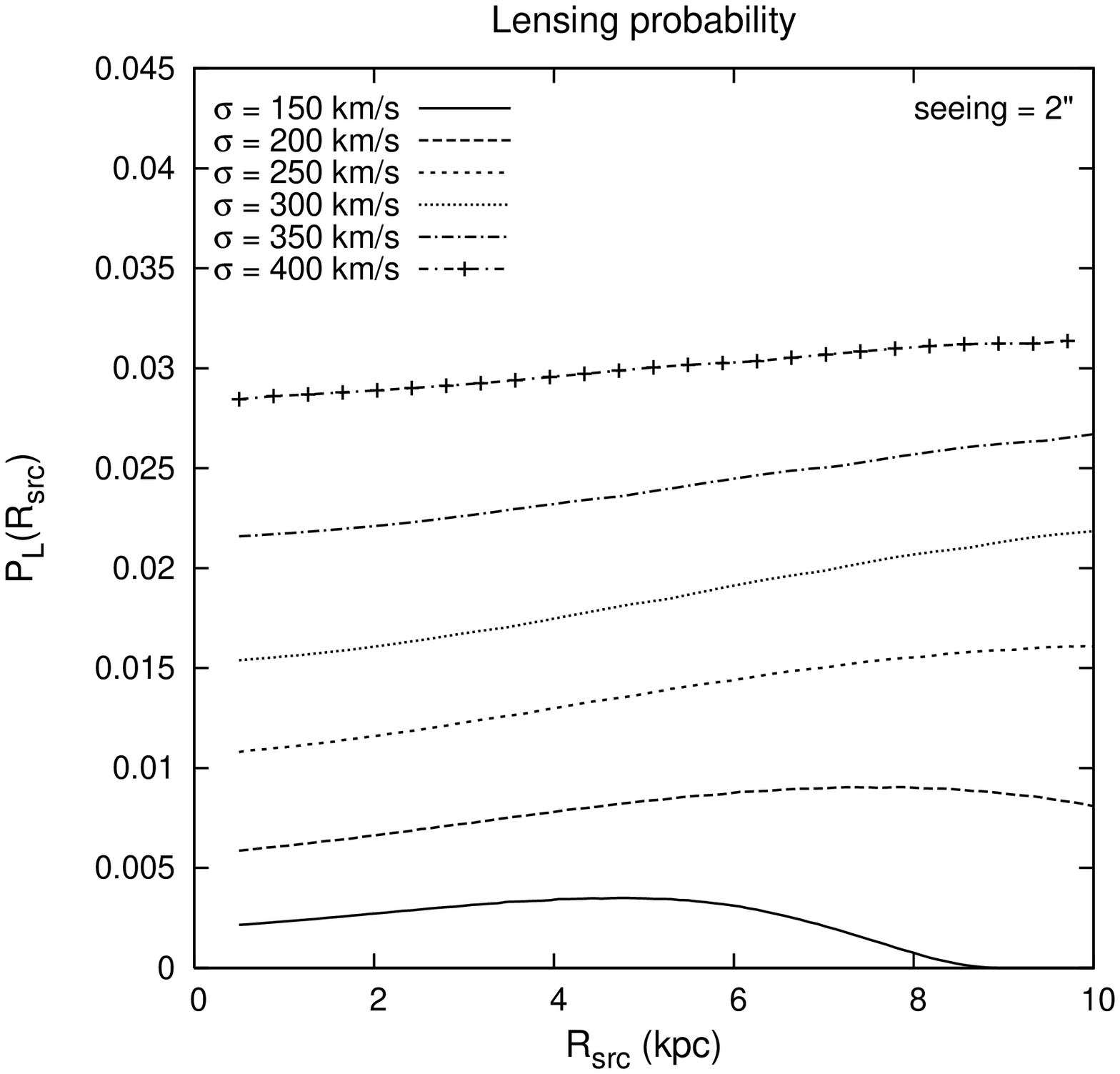}
  \includegraphics[width=0.32\textwidth]{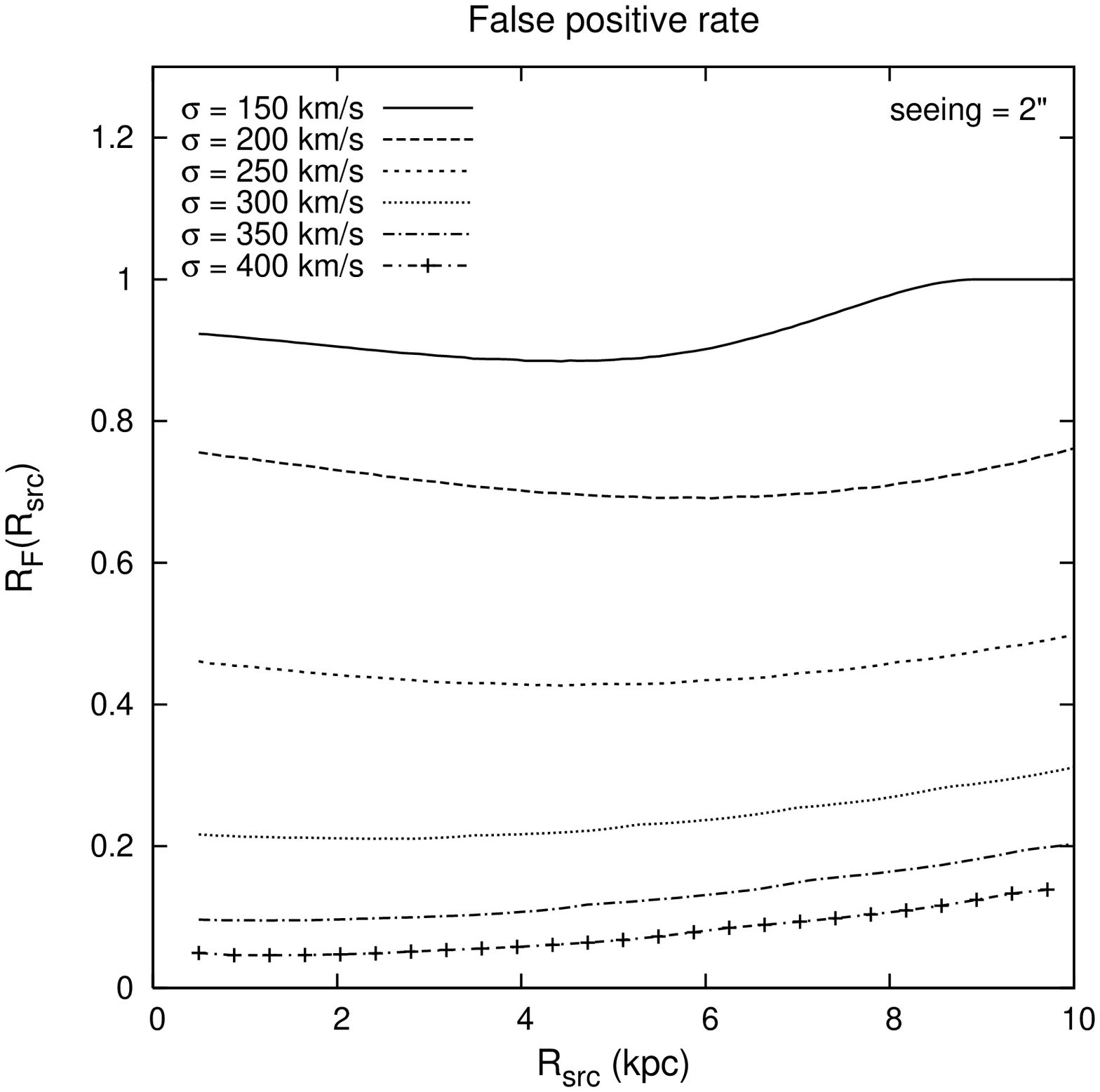}
  }
  \centerline{
  \includegraphics[width=0.32\textwidth]{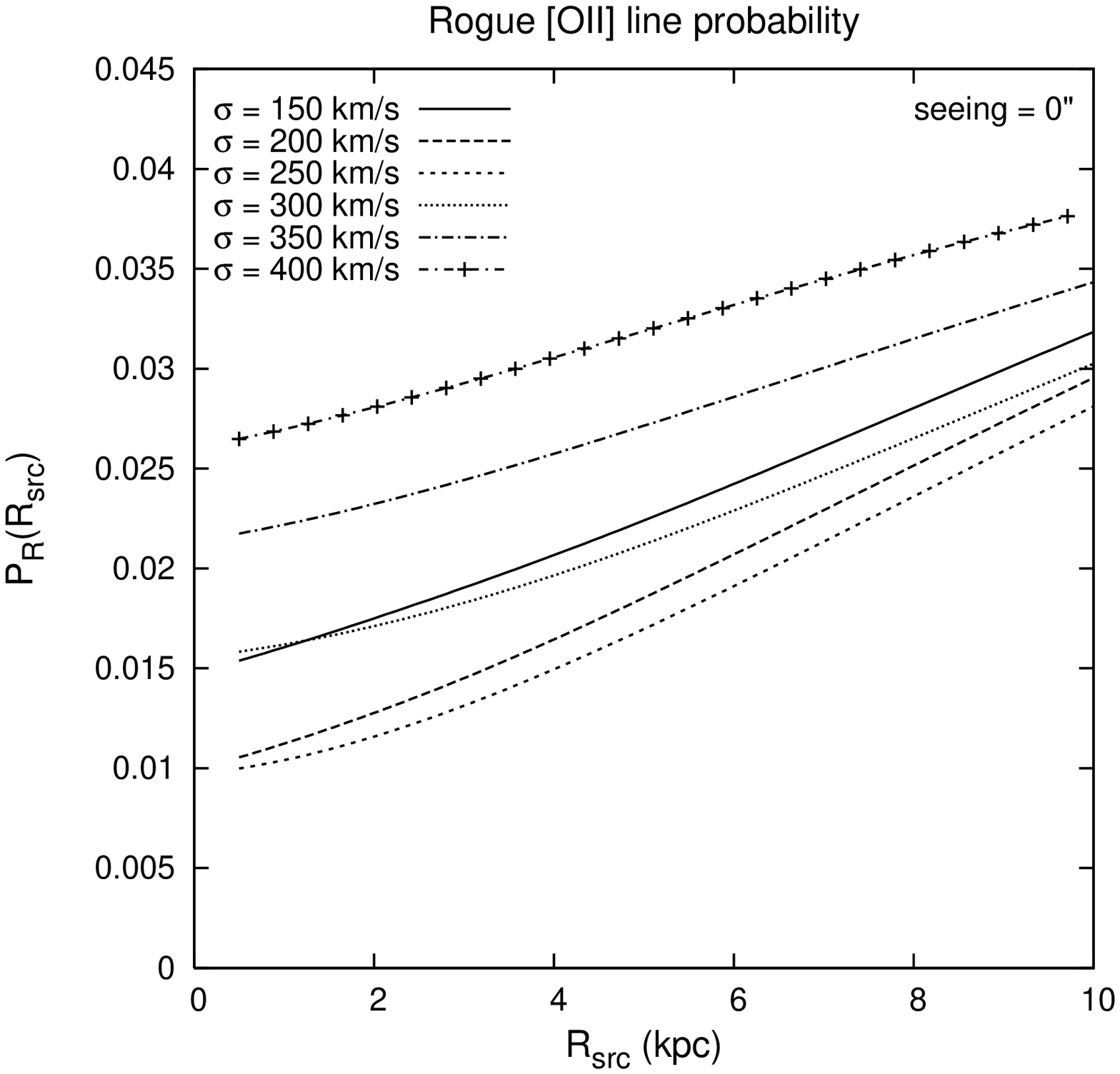}
  \includegraphics[width=0.32\textwidth]{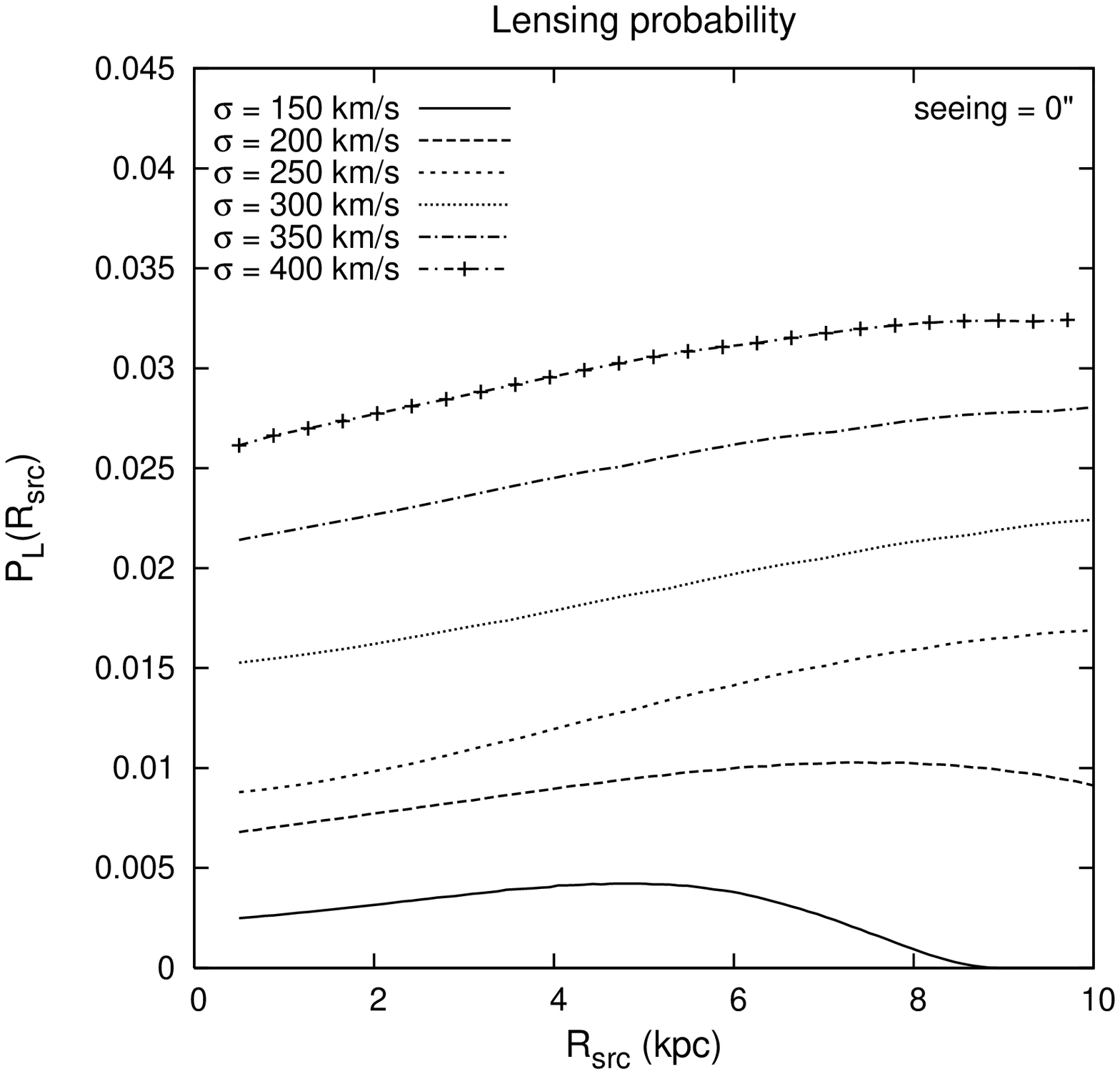}
  \includegraphics[width=0.32\textwidth]{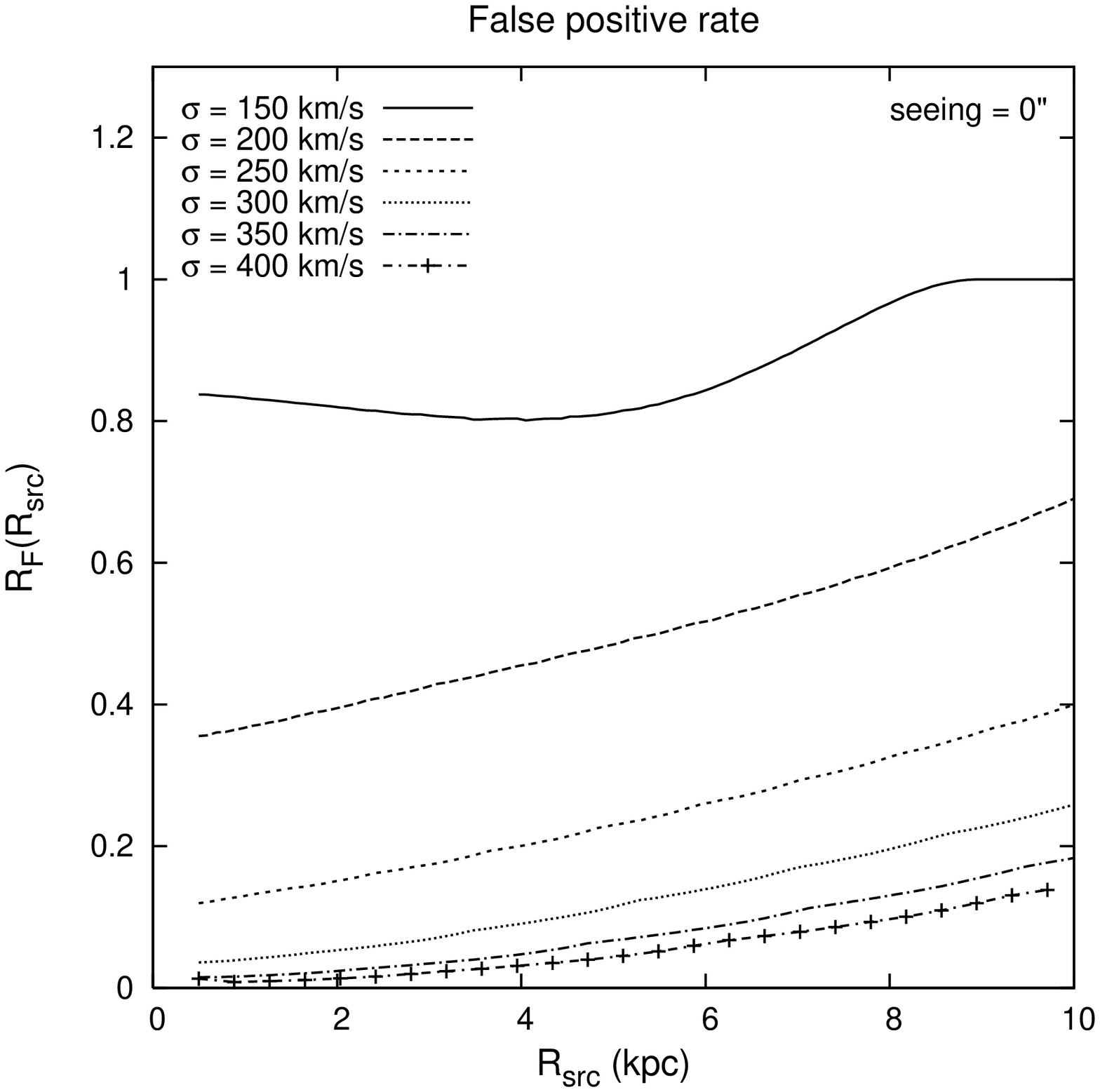}
  }
\caption{
Total rogue line probability ($P_R$, left), lensing probability
($P_L$, middle), and false positive rate ($R_F$, right), as a function
of the size of the source.  We take as fiducial parameters
$(z_l,\sigma,e,\gamma,\phi_{\gamma}) = (0.2,250\mbox{ km/s},0,0,0)$.  
In the top row we vary the lens redshift, in the middle row we vary
the velocity dispersion, and in the bottom row we vary the velocity dispersion 
without including the effects of seeing.  The spectroscopic fiber cut produces a
non-monotonic dependence of $P_R$ on $\sigma$.  This behavior also depends on 
the seeing.  The lensing probability $P_L$ turns over at large $\Rsrc$ 
(most visible in the $\sigma = 150$ km/s curve) due to the magnification cut.  
The false positive rate also exhibits complex behavior with $\Rsrc$ and is very 
sensitive to both $\sigma$ and $z_l$.
}
  \label{fig:varparams}
  \end{center}
\end{figure*}

We first seek to understand how the various probabilities depend
on the lens galaxy properties.  \reffig{varparams} shows the rogue line
probability, lensing probability, and false positive rate as a
function of source size, for different values of the lens redshift $z_l$
and velocity dispersion $\sigma$.  The total probability $P_R$ for
detecting a rogue line in an LRG spectrum is at the level of 2--3\%
(consistent with our estimate above), and depends moderately on
lens redshift.  Once a rogue line is detected, the probability that the
line corresponds to a source that is significantly distorted decreases
significantly with lens redshift: the false positive rate is only
$R_F \sim 35\%$ at $z_l=0.15$, compared with $R_F \sim 80\%$ at
$z_l=0.45$.  The trend is not surprising because there is a finite
upper limit on the source redshift, and the Einstein radius shrinks
as the lens moves out in redshift ($b \propto D_{ls}$).

As expected, \reffig{varparams} shows that the lensing probability 
depends strongly on $\sigma$: $P_L$ varies by almost a factor of
10 over the range 150 km/s $< \sigma <$ 400 km/s.  For most values
of $\sigma$, the lensing probability increases with source size
over the range $0 < \Rsrc < 10$ kpc.  However, for $\sigma=150$ km/s
the curve reaches a peak at $\Rsrc \approx 5$ kpc and then turns over.
We attribute this to finite source effects: at $\sigma = 150$ km/s
the lens is simply not strong enough to significantly perturb a
large source, ($\Rein = 0.48$ arcsec for $z_s = 0.9$).  In fact, all of the 
$P_L$ curves would turn over if we went to large enough source size.

We also find the surprising result that the rogue line probability
does not increase monotonically with $\sigma$.  With 2 arcsec seeing, the
ordering of the curves in increasing $P_R$ is $\sigma=300$, 250, 350,
200, 150, and 400 km/s (for all source sizes).  In other words, the
total rogue line probability for $\sigma=150$ km/s exceeds that for
all other cases except $\sigma=400$ km/s.  We attribute this to the
finite size of the spectroscopic fiber.  When $\sigma$ is small,
the Einstein cone is small and most sources inside the fiber are
not lensed (indeed, the false positive rate is high).  As $\sigma$
increases, lensing can push some of the light outside the fiber
(see \reffig{fibcut}), reducing the fiber flux and hence making
the rogue line harder to detect.  However, if the line is detected
the chance that it corresponds to a lens is high.  As $\sigma$
increases still further, the Einstein cone begins to fill the fiber
(the false positive rate becomes very low), and the rogue line
probability increases with $\sigma$ just like the lensing
probability.

The total rogue line probability, and its non-monotonic dependence on
$\sigma$, are sensitive to seeing.  Eliminating seeing (bottom row of
\reffig{varparams}) changes the ordering of the $P_R$ curves with
different values of $\sigma$; in particular, it shifts the case with
the lowest rogue line probability to lower $\sigma$.  By contrast,
seeing has less effect on the total lensing probability.  These
features can again be understood in terms of fiber effects.  When
$\sigma$ is low, the Einstein radius is small and most of the ``lens''
configurations lie well within the spectroscopic fiber, so seeing has
little effect on the fiber flux and hence on $P_L$.  Seeing can pull
flux into the fiber from sources that lie outside, but these are
predominantly non-lens configurations.  The net result is that seeing
increases the total rogue line probability, mainly by adding false
positives.  When $\sigma$ is high, by contrast, the fiber is mostly
filled with ``lens'' configurations, so the false positive rate is low
both with and without seeing.

Though not shown here, we have also studied the effects of varying
ellipticity and shear parameters.  We find no significant change in
the lensing probability due to $e$ and $\gamma$, in contrast to
point source lensing statistics \citep{huterer}.  The difference is
presumably related to the different statistical question (i.e., the
probability that a galaxy is a lens rather than the probability
that a source is lensed), and to effects like the fiber cut that
are specific to spectroscopic surveys.

\subsection{Dependence on Survey Parameters}

\begin{figure*}
  \begin{center}
  \centerline{
  \includegraphics[width=0.32\textwidth]{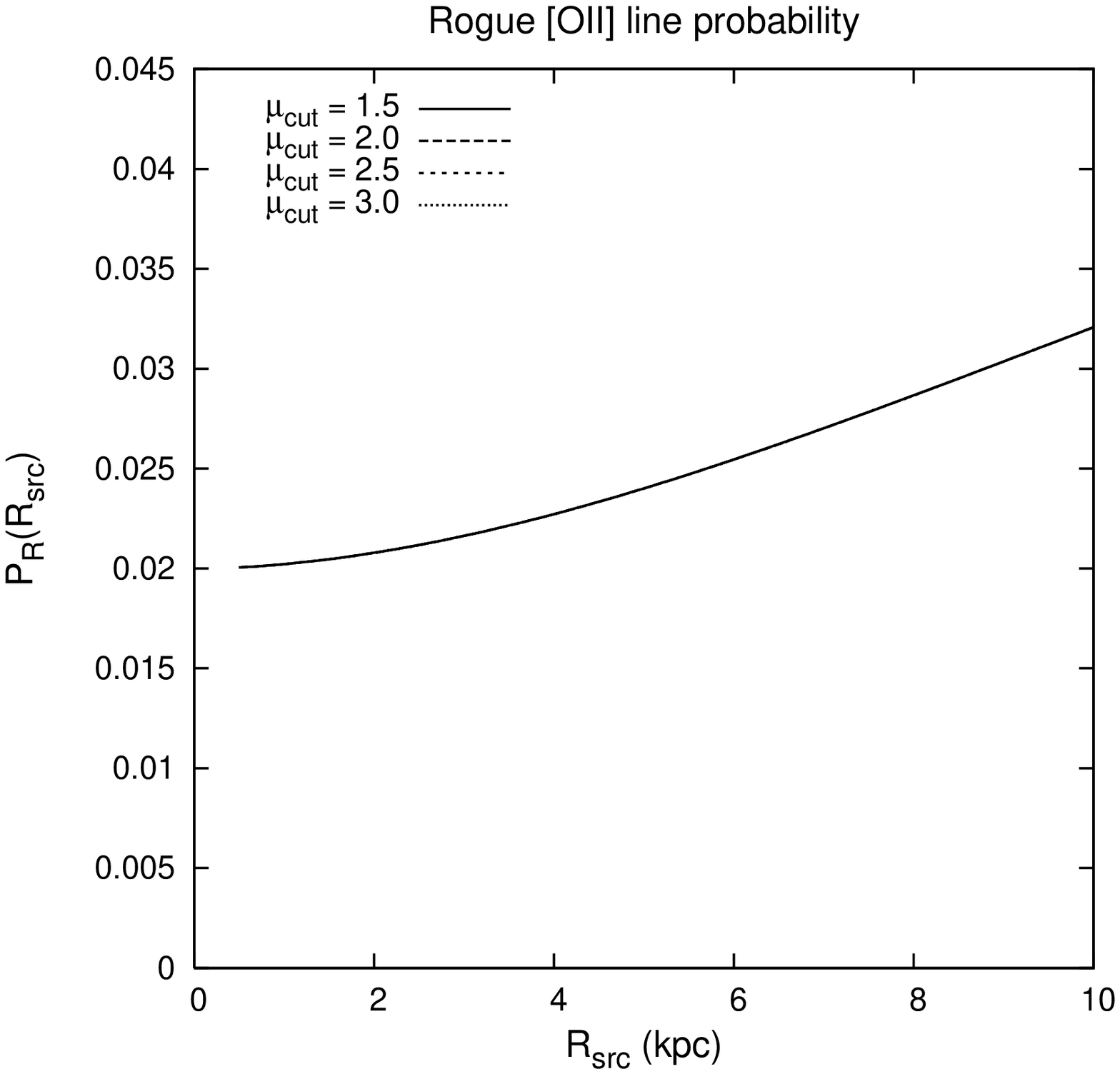}
  \includegraphics[width=0.32\textwidth]{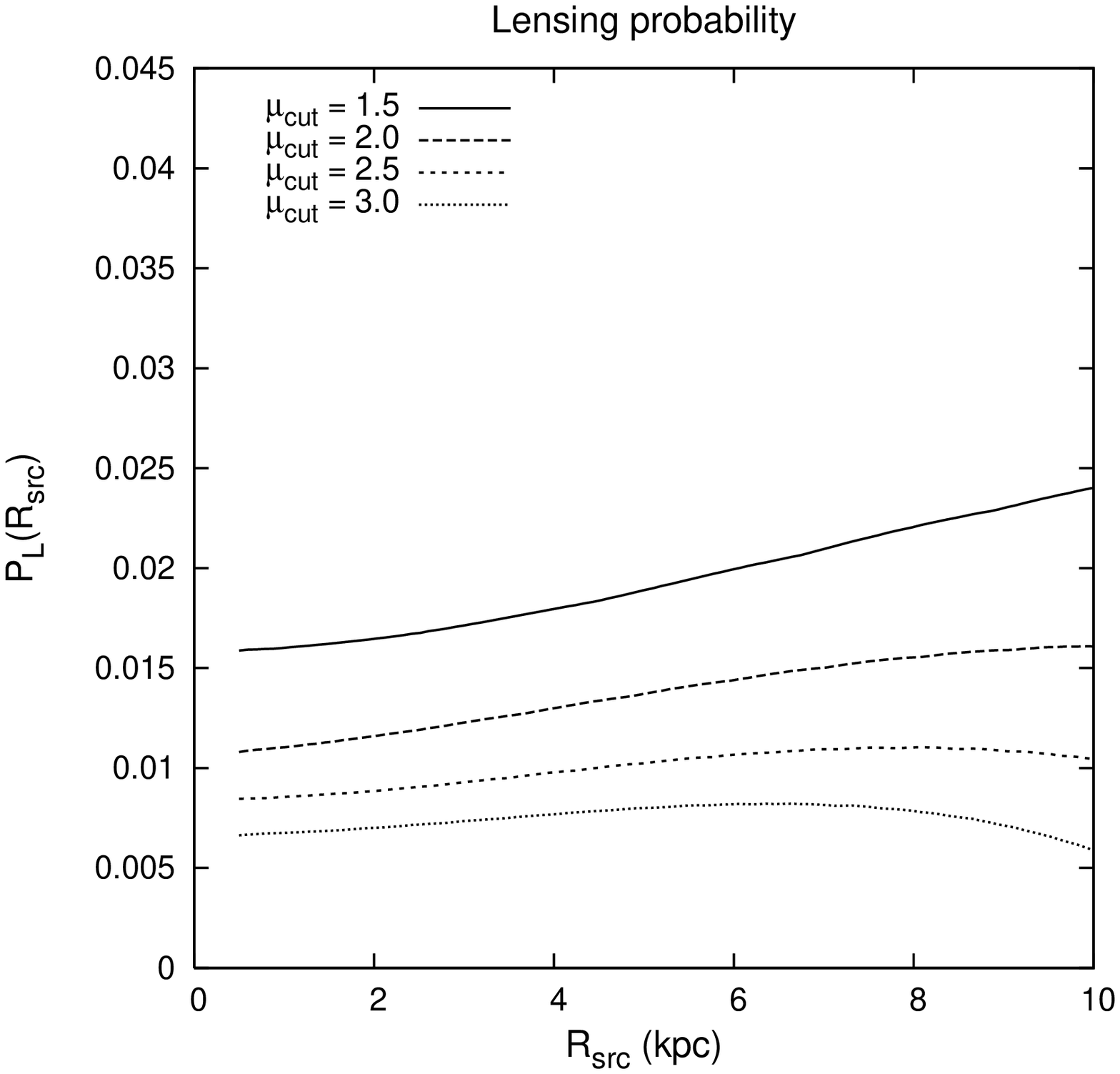}
  \includegraphics[width=0.32\textwidth]{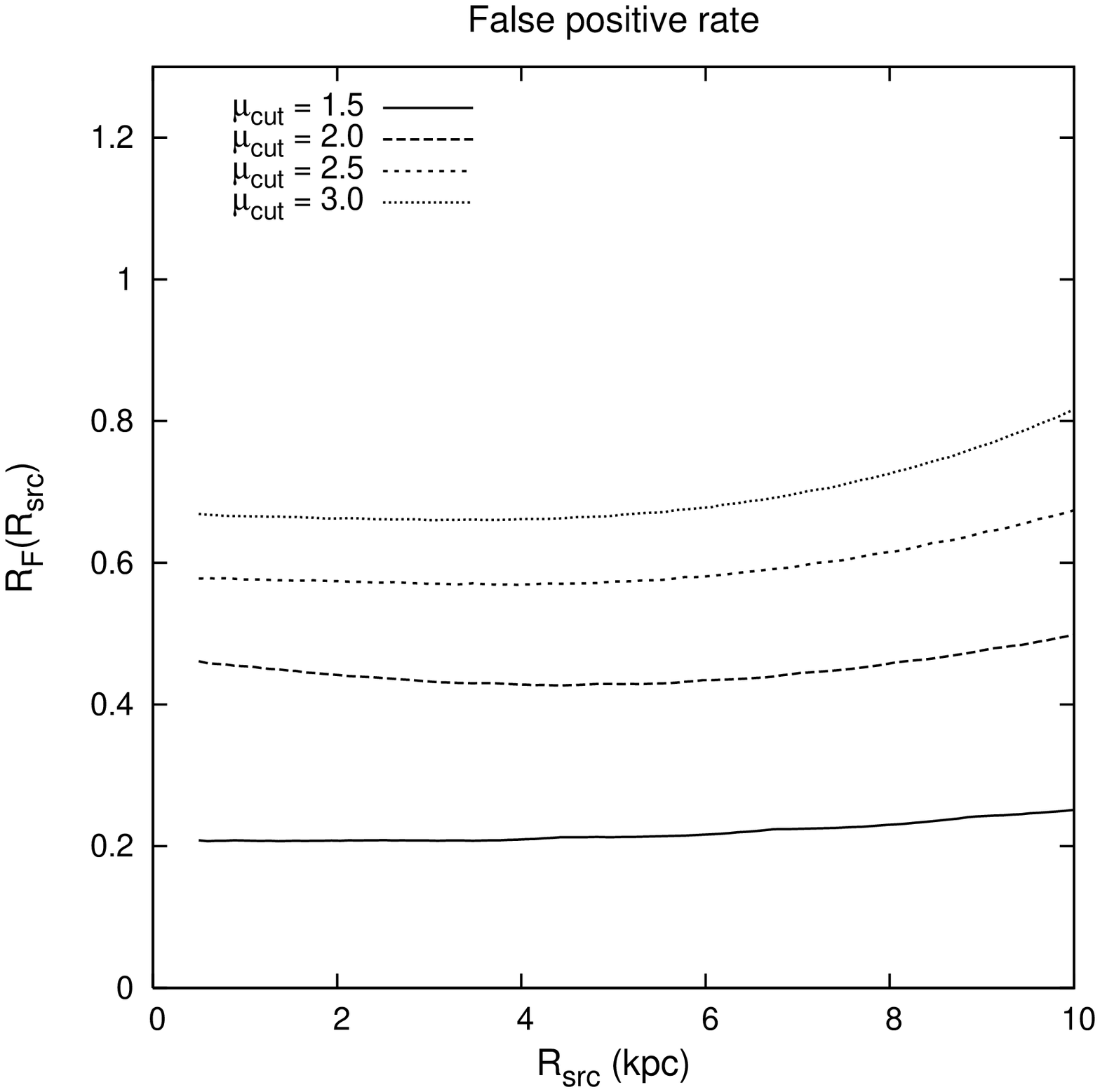}
  }
\caption{
Same as \reffig{varparams} except for varying $\mu_{\rm cut}$.
The total number of rogue lines remains unchanged (of course), but the 
lensing probability $P_L$ varies by a factor 2.4 at small source size and 4.1 at 
large source size.  As expected, the false positive rate $R_F$ is sensitively 
dependent on the definition of a lens, $\mu_{\rm cut}$.
}
  \label{fig:mucut}
  \end{center}
\end{figure*}

It is important to understand the dependence of the probabilities on our choice
of the definition of a lens (the magnification cut $\mucut$), and on the survey
noise floor $S_0$.  \reffig{mucut} shows $P_R$, $P_L$, and $R_F$ for
$1.5 \leq \mucut \leq 3.0$.  Since $\mucut$ simply indicates whether a source
is classified as lensed or not, the rogue line probability does not change.
However, the strong lensing probability $P_L$ \emph{does} depend on $\mucut$.
For $\Rsrc = 0.5$ kpc, increasing $\mucut$ from 1.5 to 3.0 decreases the
lensing probability by a factor of 2.4.  At $\Rsrc = 10$ kpc, the change is a
factor of 4.1.  The dependence on source size can be understood in terms of
the magnification regions in the source plane.  As $\mucut$ is increased,
the total area of the source plane with $\mu > \mucut$ decreases more slowly
for small sources than for larger sources.

If we double the noise floor $S_0$, we find that the rogue line probability
is decreased by a factor of 0.65 for $\Rsrc=0.5$ kpc, and 0.63 for
$\Rsrc=10$ kpc.  This change is mainly caused by the number of sources brighter
than the flux limit.  Since $n_s \propto \Gamma[1+\alpha,L_0/\mu L_*]$ and
$L_0 \propto S_0$, doubling the noise floor changes the number of detectable
sources by a factor of
$\sim \Gamma[1+\alpha,2 L_0/\mu L_*]/\Gamma[1+\alpha,L_0/\mu L_*]$.  This
works out to be a factor of 0.6--0.8 for typical $L_0$ and $\mu$ values.
This simple estimate agrees quite well with our full calculations despite
ignoring complicated seeing effects, magnification effects, and the redshift 
dependence of $L_0$.

\subsection{Total Probabilities and Higher Order Statistics}
\label{sec:totprob}

\begin{figure}
  \begin{center}
  \includegraphics[width=0.45\textwidth]{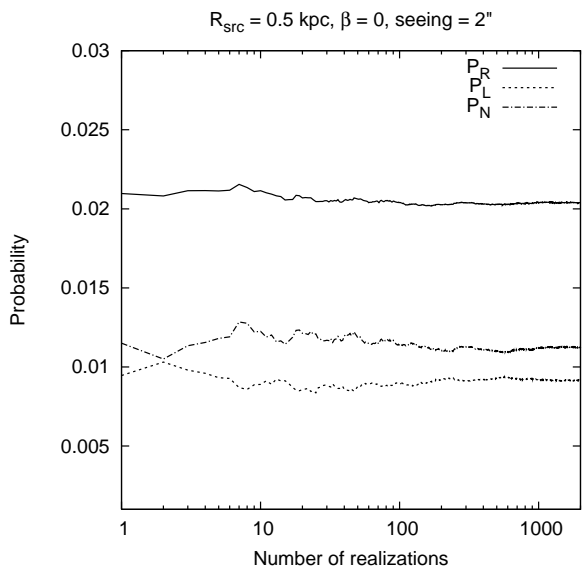}
\caption{
Total detection probabilities as a function of the
number of LRG realizations.  In this convergence test we fix the
source size to be $\Rsrc = 0.5$ kpc and we use a non-evolving source
luminosity function ($\beta=0$).  The probabilities converge quickly,
indicating that 800 realizations is more than adequate to yield
accurate statistics.
}
  \label{fig:converge-test}
  \end{center}
\end{figure}

\begin{figure*}
  \begin{center}
  \centerline{
  \includegraphics[width=0.32\textwidth]{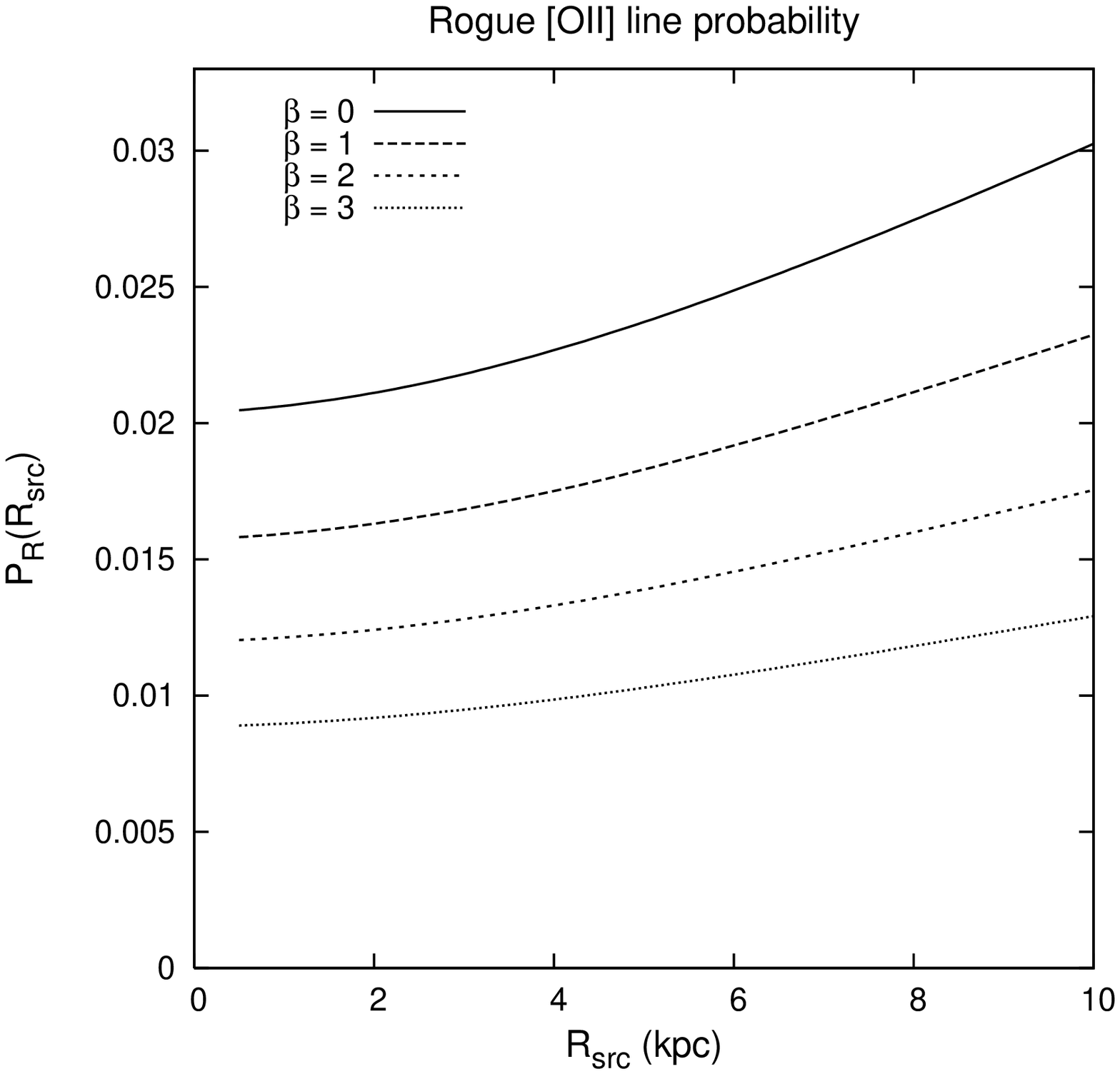}
  \includegraphics[width=0.32\textwidth]{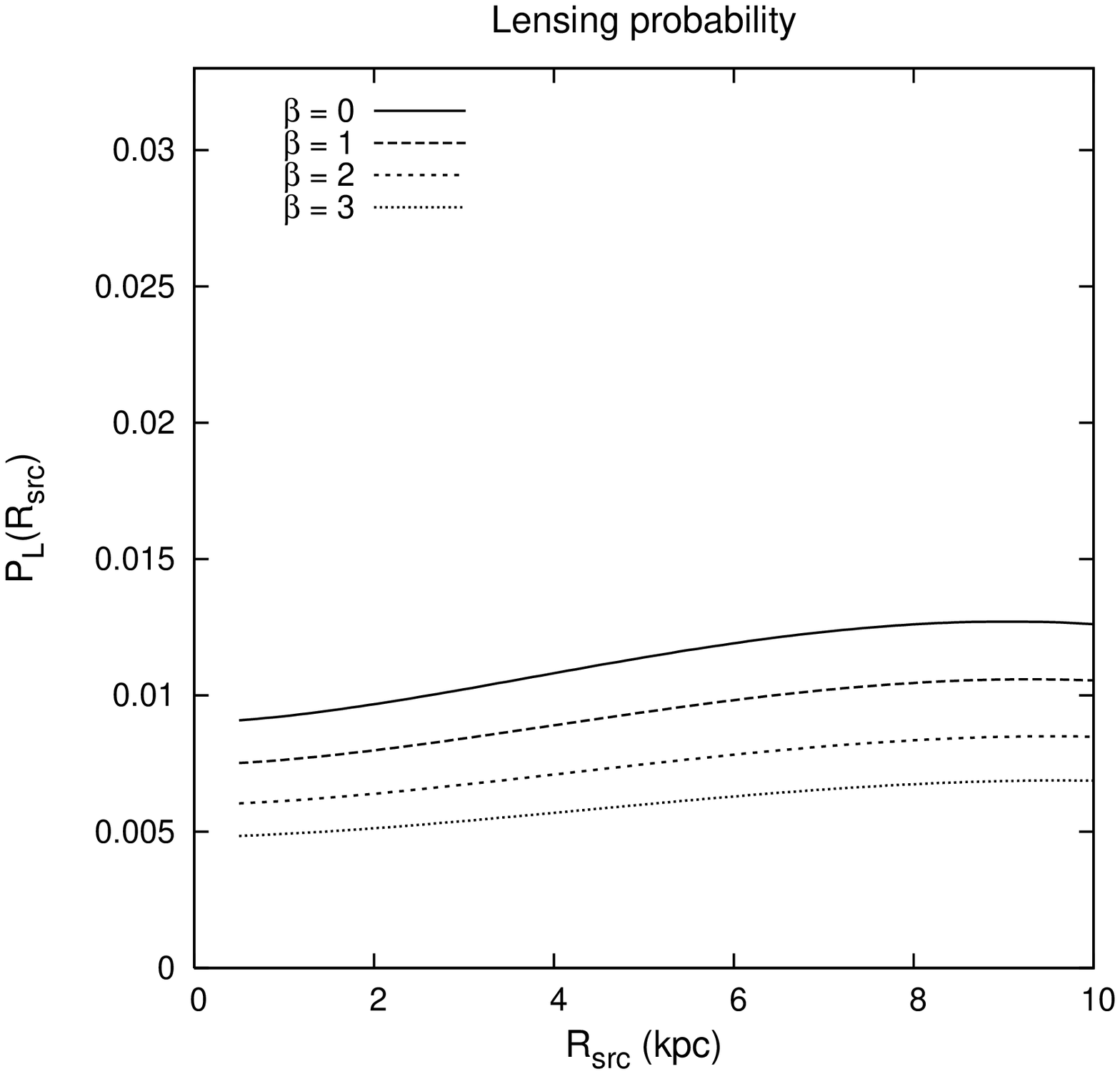}
  \includegraphics[width=0.32\textwidth]{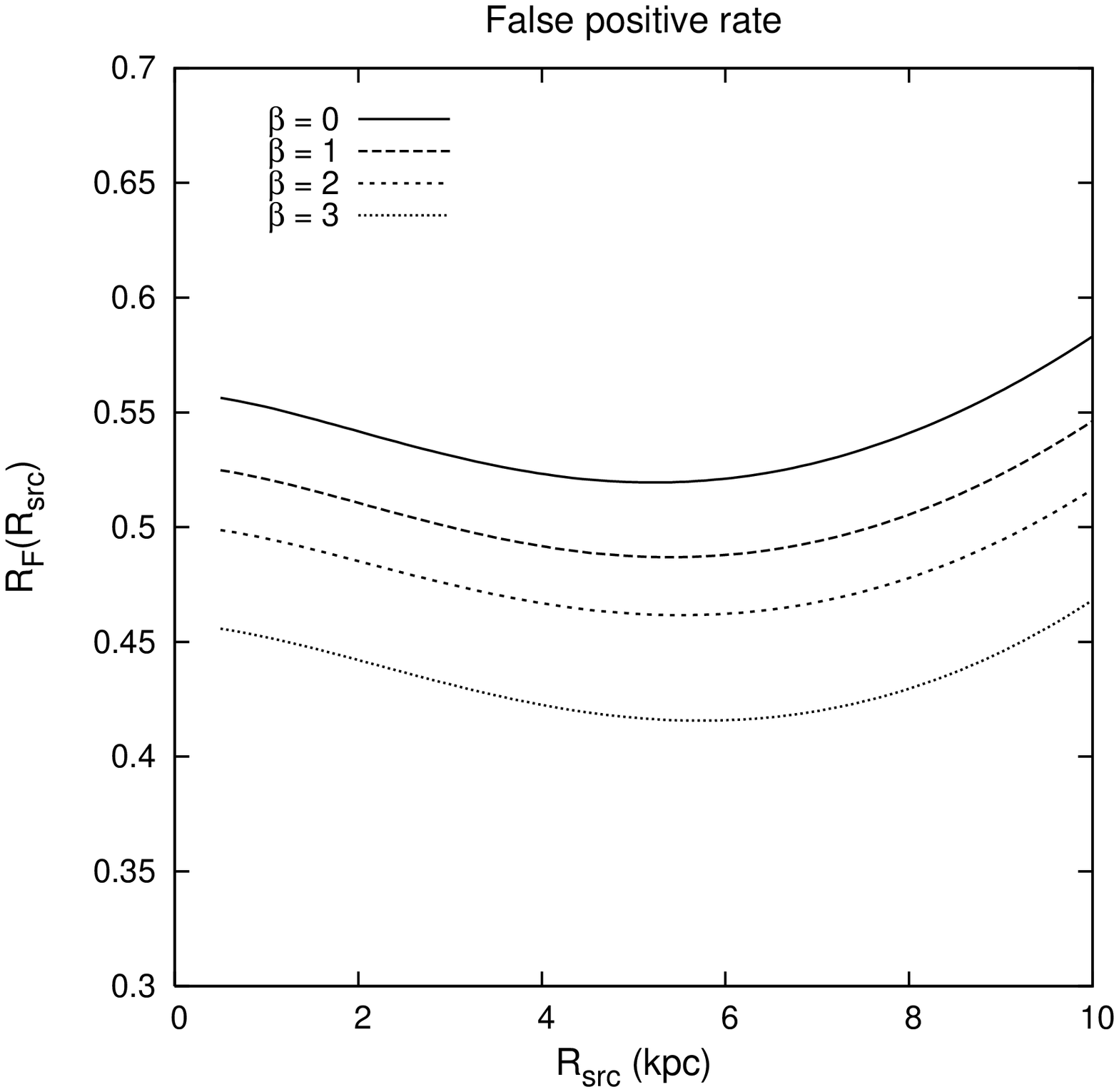}
  }
\caption{
Total detection probabilities after averaging over 800 sets of LRG
parameters, for four different models of LF evolution.  Given these
results, we expect that the initial SLACS sample of 50,996 LRGs should contain
$\sim$460--1,530 galaxies with rogue emission lines in their spectra, and
$\sim$250--640 should reveal strong lensing features in follow-up
observations.
}
  \label{fig:finalprob}
  \end{center}
\end{figure*}

\begin{figure*}
  \begin{center}
  \centerline{
  \includegraphics[width=0.32\textwidth]{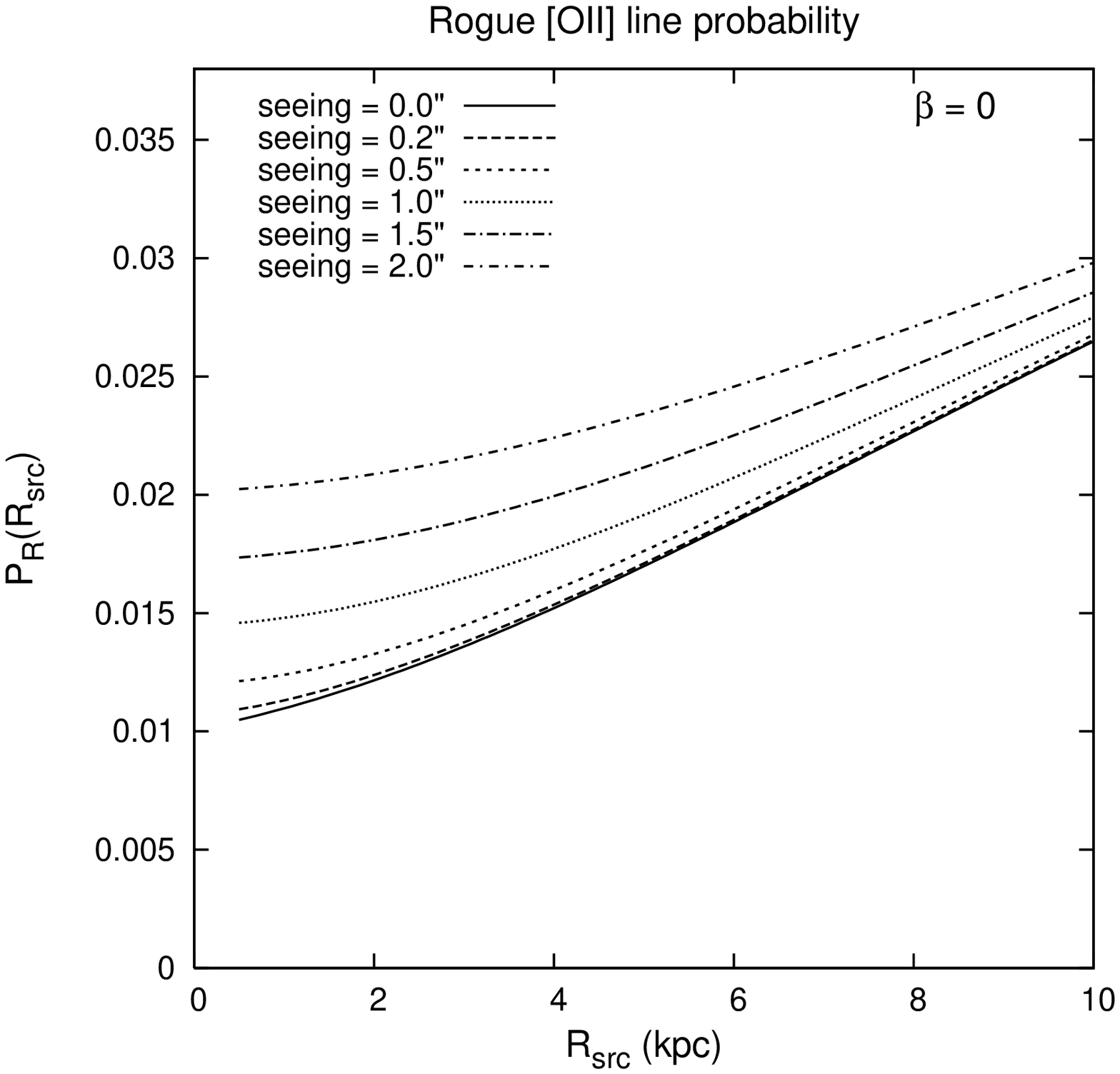}
  \includegraphics[width=0.32\textwidth]{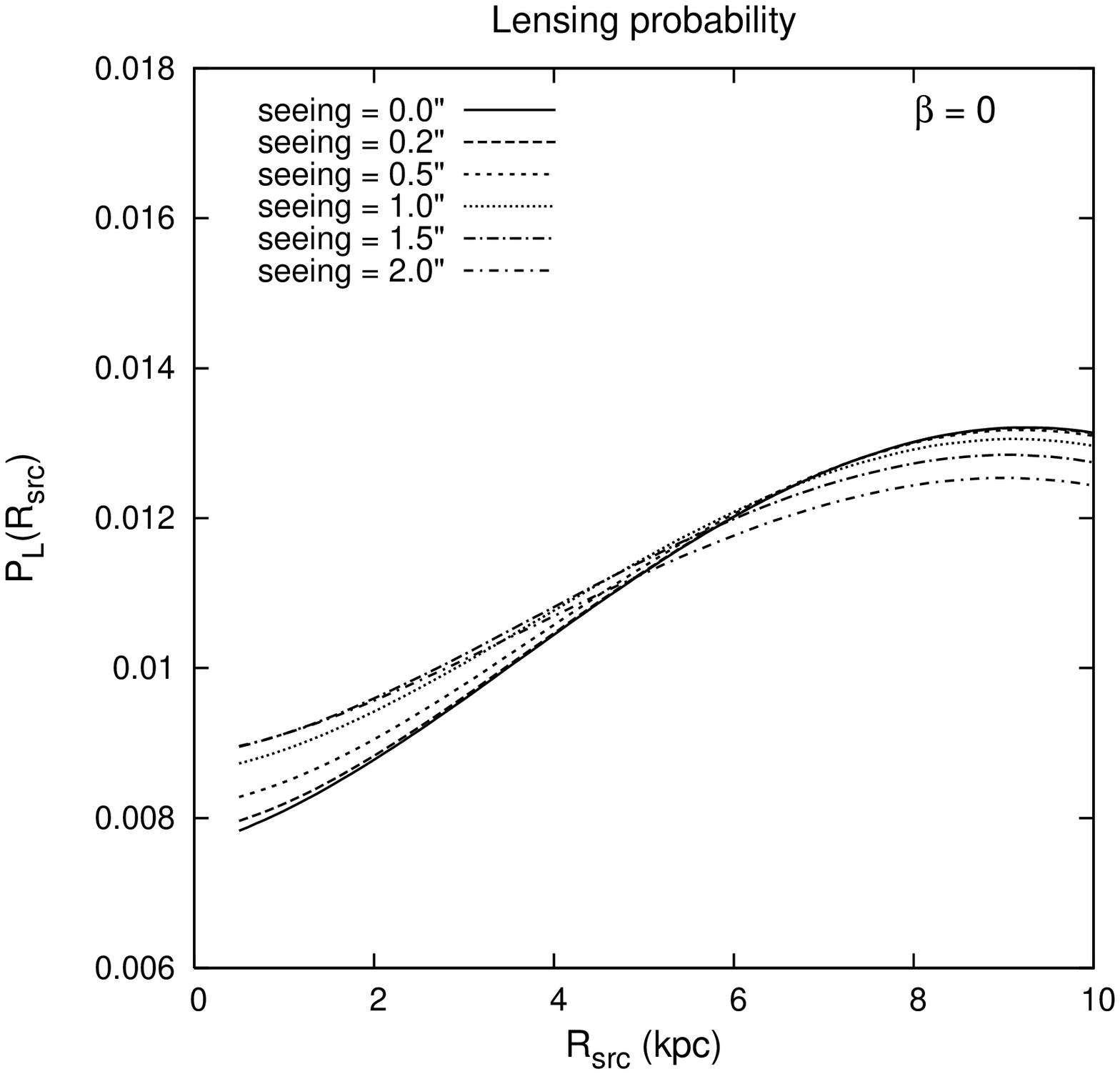}
  \includegraphics[width=0.32\textwidth]{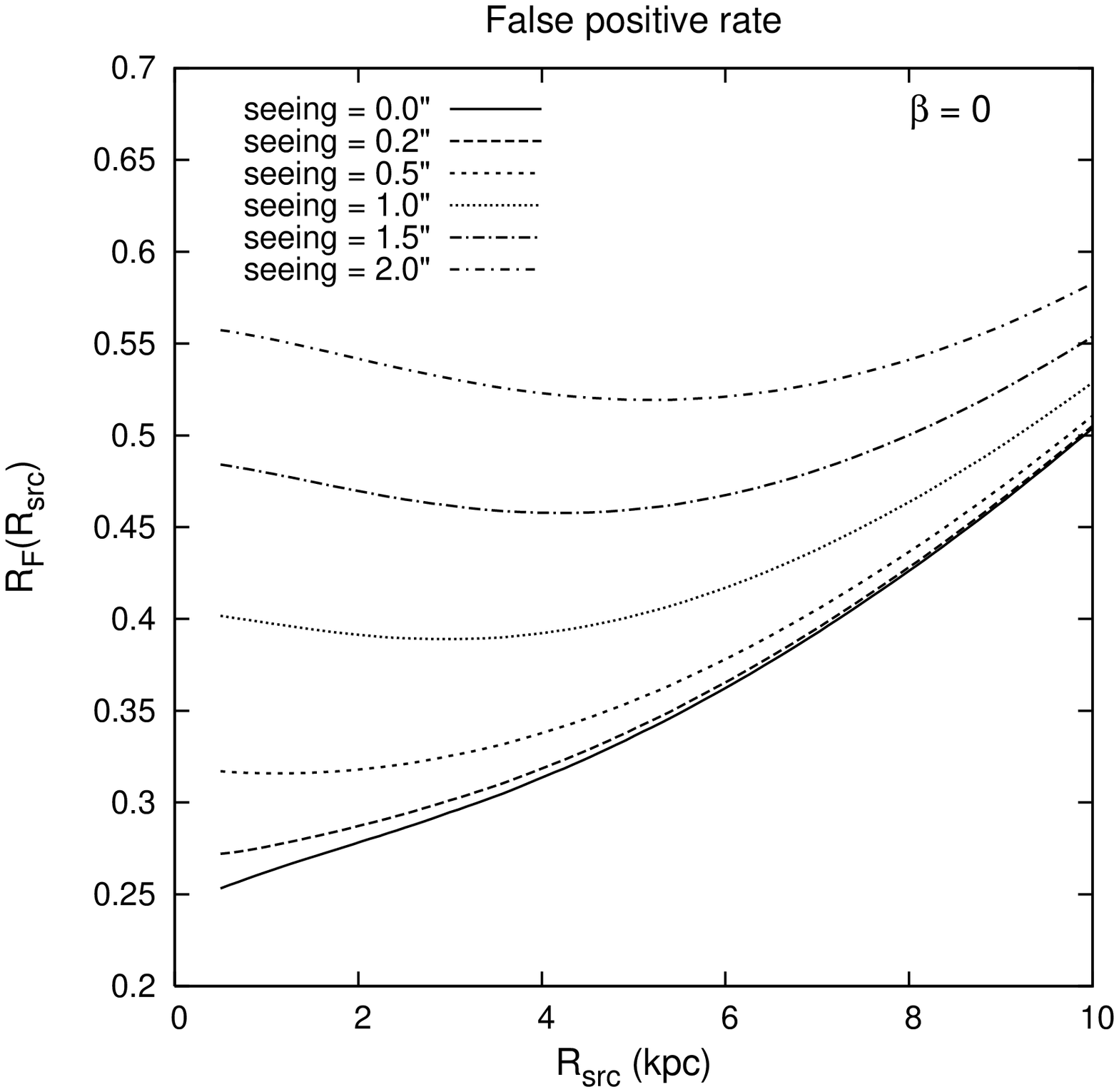}
  }
\caption{
Total detection probabilities as a function of source size and the FWHM of the 
Gaussian seeing model ($\beta = 0$).  Seeing effects boost the lensing 
probability at small $\Rsrc$ but suppress the lensing probability at large 
$\Rsrc$.  The total rogue line detection probability is boosted for all source 
sizes.  The fraction of non-lenses is much more sensitive to seeing effects 
at small $\Rsrc$ leading to the minimum in the false positive rate in 
\reffig{finalprob}.
}
  \label{fig:vary-seeing}
  \end{center}
\end{figure*}

We are now ready to compute overall probabilities by averaging over an
appropriate sample of galaxies (see eq.\ \ref{eq:totprob}).  First we need to
consider how many galaxies we need to include to obtain accurate
statistics.  \reffig{converge-test} shows the three probabilities as a function 
of the number of LRG realizations (for $\Rsrc=0.5$ kpc and $\beta=0$, 
meaning no evolution in the source [\ion{O}{2}] luminosity function).  All
three probabilities converge rather quickly, indicating that our
fiducial sample size of 800 LRGs is sufficient.

\reffig{finalprob} shows the total probability for detecting rogue emission 
lines in LRG spectra, the probability of identifying lensing in follow-up 
observations, and the SLACS false-positive rate all as a function of source size 
averaged over our full sample of 800 galaxies.

At this point we also examine the effects of evolution of the
source [\ion{O}{2}] luminosity function.  We consider four models,
ranging from no evolution ($\beta=0$) to a strong increase in the
number of [\ion{O}{2}] emitters in the past ($\beta=3$).  A value of
$\beta \approx 2$--3 for $z<1$ is preferred by \citet{glaze} based
on SDSS derived star formation histories.  LF evolution tends to
decrease the rogue line and lensing probabilities.  The explanation
is that we fix the total number density of sources in the range
$0.35 \leq z \leq 1.5$ from the observations by \citet{hogg}, so
increasing $\beta$ shifts a higher fraction of those sources to
higher redshifts, and therefore \emph{decreases} the number of
sources below the SLACS upper limit $z_{s,{\rm max}} = 0.9$.

For a source size of 0.5 kpc, we find the total probability for rogue line 
detection varies from 2.0\% for source LF's with no evolution ($\beta=0$) to 
0.9\% for LF's with strong evolution ($\beta=3$).  With $\Rsrc=10$ kpc, the 
probabilities are 3.0\% ($\beta=0$) and 1.3\% ($\beta=3$).  The implication is 
that in the original \citet{bolton04} sample of 50,996 LRG spectra, 
$\sim$460--1,530 should contain rogue [\ion{O}{2}] emission lines in their 
spectra above a threshold of $6 \times 10^{-17}$ erg/s/cm$^2$.  Of those, 
$\sim$250--640 should show significant strong lensing features when imaged with 
high spatial resolution.  The remaining systems are false positive detections in 
which there is a background galaxy that is not significantly perturbed by 
lensing effects.  The total false positive rate $R_F(\Rsrc)$ also varies 
significantly with with source size and $\beta$ but has typical values $R_F \sim$ 
50\% (see \reffig{finalprob}).  It is important to note that while the 
breakdown into lenses and false positives depends on our choice of the 
magnification cut ($\mu_{\rm cut} = 2.0$), the total number of rogue lines is a 
robust prediction.

While our fiducial results have been computed for 2 arcsec seeing, it is
instructive to consider how seeing affects our result.  \reffig{vary-seeing} 
shows the three probabilities as a function of source size for various
values of the seeing.  The effect of seeing is to introduce a ``tilt'' to
the lensing probability curve, giving a moderate boost at small source
size and reduction at large source size.  Seeing increases the total
rogue line probability, especially at small source size.  The
implication is that seeing enhances the number of detected rogue lines that
do not correspond to lens systems at all source sizes but most
dramatically at small $\Rsrc$.  Indeed, the false positive rate
increases monotonically with $\Rsrc$ when seeing is unimportant (the
FHWM is small compared with the size of the spectroscopic fiber), while it
flattens out and develops a minimum near $\Rsrc \approx 5.5$ kpc when
the seeing is important.

Although there is significant uncertainty due to source size and LF 
evolution effects, our lensing estimates are clearly higher than the $\sim$150 
rogue lines and $\sim$60 lens systems found in the initial SLACS sample.  The 
most likely explanation for this discrepancy is that the SLACS selection 
criteria require two \emph{additional} emission features ($H\beta$, [\ion{O}{3}] 
4959, or [\ion{O}{3}] 5007) besides the primary [\ion{O}{2}] 3727 line.  The 
presence of these additional features was required to substantially reduce
the number of false positives, but may have eliminated many
real lens systems as well.  Incorporating multiple emission
line statistics into our calculations would require knowledge of
the joint probability distribution between [\ion{O}{2}] and
secondary line luminosities and is beyond the scope of the
present study.  Furthermore, secondary lines tend to appear in the 
$\sim$7000--9000 \AA\ region of the spectrum where sky noise is more problematic;
many spectra with secondary features buried in the noise may
have been rejected as targets.  In any event, we predict that there is a large 
number of g-g lenses waiting to be discovered in the SDSS spectroscopic data.

\begin{figure*}
  \begin{center}
  \centerline{
  \includegraphics[width=0.4\textwidth]{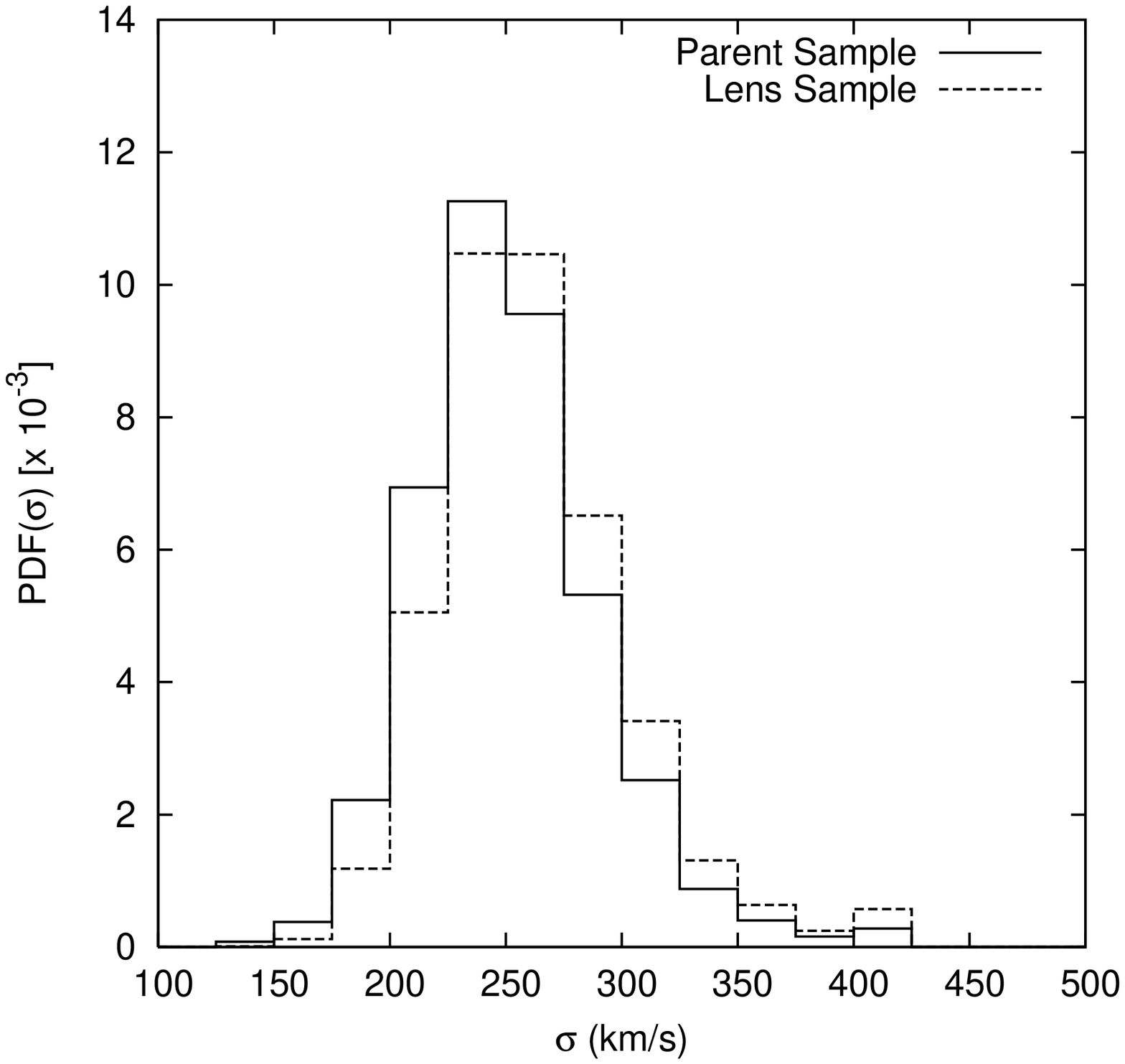}
  \includegraphics[width=0.4\textwidth]{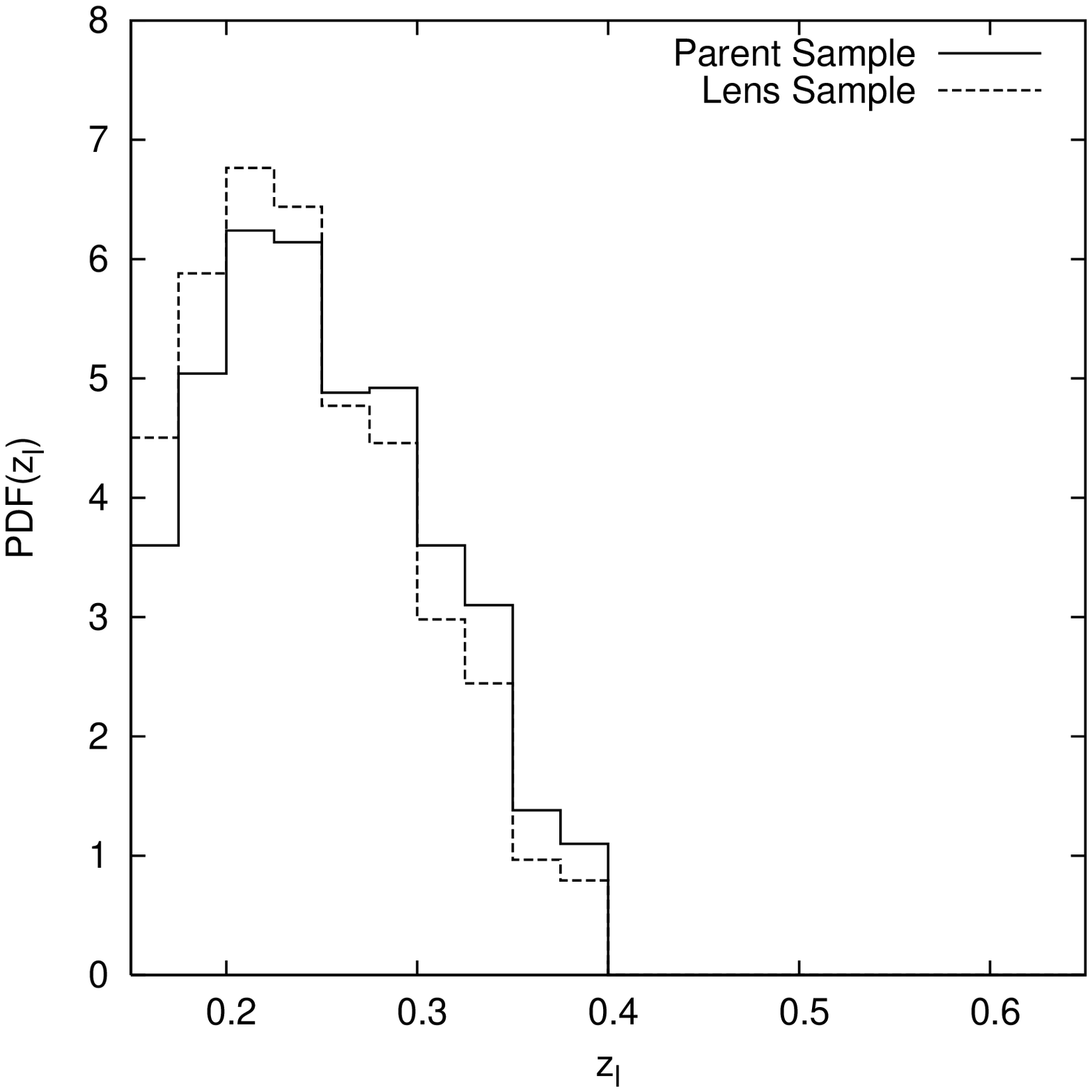}
  }
  \caption{
Histograms of velocity dispersion (left) and redshift (right), for
the LRG parent sample (solid line) and the subsample that represent
lenses (dashed line, $\Rsrc = 0.5$ kpc, $\beta = 0$, 
seeing=2\arcsec{}).  All histograms are normalized to unit area.
The lens galaxies are biased towards higher velocity dispersion and
lower redshift.  A KS test comparing the distributions of the velocity dispersion 
(lens redshift) yields only a $2.5\times 10^{-7}$\% (0.05\%) chance that the 
populations are drawn from the same distribution.
}
  \label{fig:histograms}
  \end{center}
\end{figure*}

Finally, it is interesting to consider how the galaxies that produce
g-g lenses may form a biased subset of all LRGs.  \reffig{histograms}
shows the distributions of redshift and velocity dispersion for our
parent population of galaxies, and for the same population with each
galaxy weighted by its lensing probability (with $\Rsrc = 0.5$ kpc, 
$\beta = 0$, seeing=2\arcsec{}).  There is clearly a bias
towards larger velocity dispersions, which is natural, but it is
weaker than in traditional quasar lens statistics because of fiber
cut effects.  There is also a bias towards lower redshifts, because
lower-redshift galaxies have larger Einstein cones (cf.\ 
\reffig{varparams}).  We test the null hypothesis that the two populations 
represent the same probability distribution with the Kolmogorav-Smirnov test 
\citep{press92}.  Comparing the distributions of the velocity dispersion 
(redshift) yields only a $2.5\times 10^{-7}$\% (0.05\%) chance that the 
populations are drawn from the same distribution.

\section{Conclusions}

We have introduced a statistical method to estimate the expected
number of strong galaxy-galaxy lens systems in a spectroscopic survey.
In the process, we have also developed a semi-analytic technique for
calculating the magnification of a finite source due to an isothermal
ellipsoid galaxy in an external shear field.  Conceptually, the most
important result is that the size of the spectroscopic fiber provides
a significant selection effect.  In our analysis of the SLACS survey,
the fiber cut yields the unexpected result that the probability of
detecting a rogue emission line is a non-monotonic function of the
velocity dispersion $\sigma$.  Since larger $\sigma$ corresponds to
a larger Einstein cone, one would naively expect the rogue line
probability to increase with $\sigma$.  However, large $\sigma$ also
corresponds to a large image separation, which can cause much of the
source flux to fall outside the spectroscopic fiber.  The situation is further 
complicated by the effects of atmospheric seeing which can add flux into the fiber 
from images outside or remove flux from the fiber from images inside.  It will be
crucial to compute the effects of the fiber cut and seeing, customized to the
parameters of the spectrograph, for any future search for g-g
lenses in large spectroscopic surveys.

The lensing probability has a fairly weak dependence on the size
of the source, but a stronger dependence on the evolution of the
source luminosity function.  Lensing introduces biases such that
lens galaxies will tend to have higher velocity dispersions and
lower redshifts compared with the parent population of galaxies.

Incorporating parameters from the SLACS survey, we estimate that
there should be $\sim$460--1,380 rogue emission lines in the sample
of 50,996 LRG spectra.  Of these, $\sim$250--640 should show clear
evidence of strong lensing in follow-up observations.  The broad range of 
probabilities is due primarily to uncertainties in the physical size of 
[\ion{O}{2}] emission regions and in the evolution of the [\ion{O}{2}] 
luminosity function.  Specifically, small sources and strong evolution yield
lower probabilities while large sources and no evolution give higher 
probabilities.  Our estimates are notably higher than the numbers actually 
observed in the SLACS survey so far.  We attribute this to their requirement 
that multiple emission lines be detected, to the large amount of sky noise in 
the long wavelength region of the spectra, and to the potential extinction of 
the [\ion{O}{2}] emission line by dust in the lens galaxies.  Our
calculations imply that there are many more galaxy-galaxy strong
lenses waiting to be found in the SDSS spectroscopic data.

While our methods have been specifically applied to the SLACS
survey, they should be applicable to all finite source lens
searches in upcoming spectroscopic surveys.  Future improvements
to our method would involve incorporating non-uniform brightness
distributions for the background source galaxies.

\acknowledgements
GD and CRK were supported in part by grant HST-AR-10668 from the Space 
Telescope Science Institute, which is operated by the Association of 
Universities for Research in Astronomy, Inc., under NASA contract NAS5-26555.

\appendix

\section{Two-Dimensional Integration}
\label{app:appendix}

Our source plane integration is carried out via a 2-D integrator that tiles the 
integration region efficiently by using a movable grid with adaptive 
resolution.  As an example to illustrate our method, we first consider the 
magnification calculation for a uniform brightness finite source (with size 
$\Rsrc$ centered at the origin) which is lensed by an isothermal ellipsoidal mass 
distribution (SIE).

Since the magnification is defined as,
\be
  \mu = \frac{\mbox{Area of the images}}{\mbox{Area of the source}},
\ee
we must find the total area of the images by integrating over the 2-D image 
plane.  We start with a single ``macro''-grid whose lower right corner is 
centered on the origin and suppose that the edge of an image passes through this 
grid as shown in Figure \ref{fig:inout}.
\bp
  \begin{center}
  \includegraphics[width=0.45\textwidth]{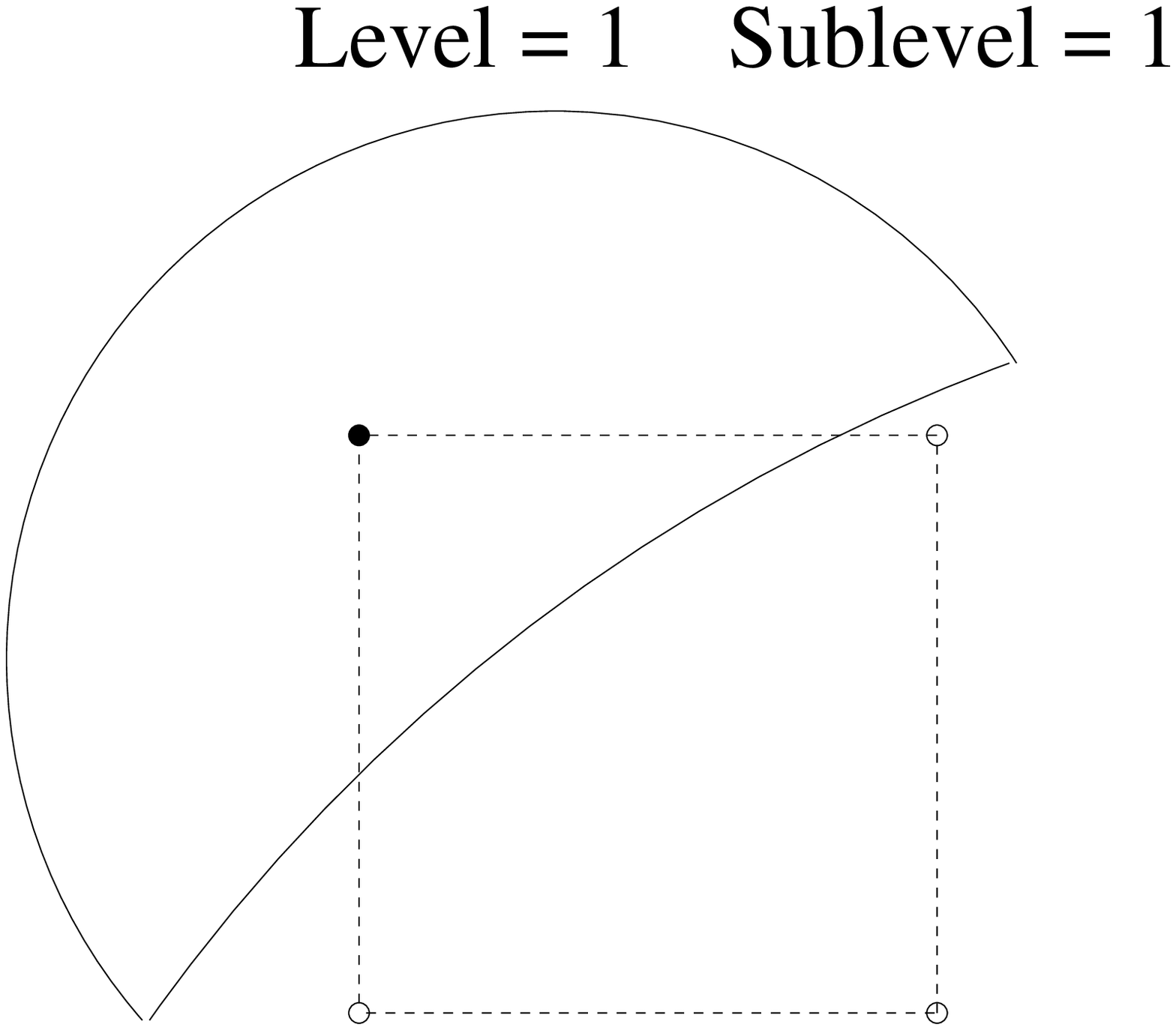}
  \includegraphics[width=0.45\textwidth]{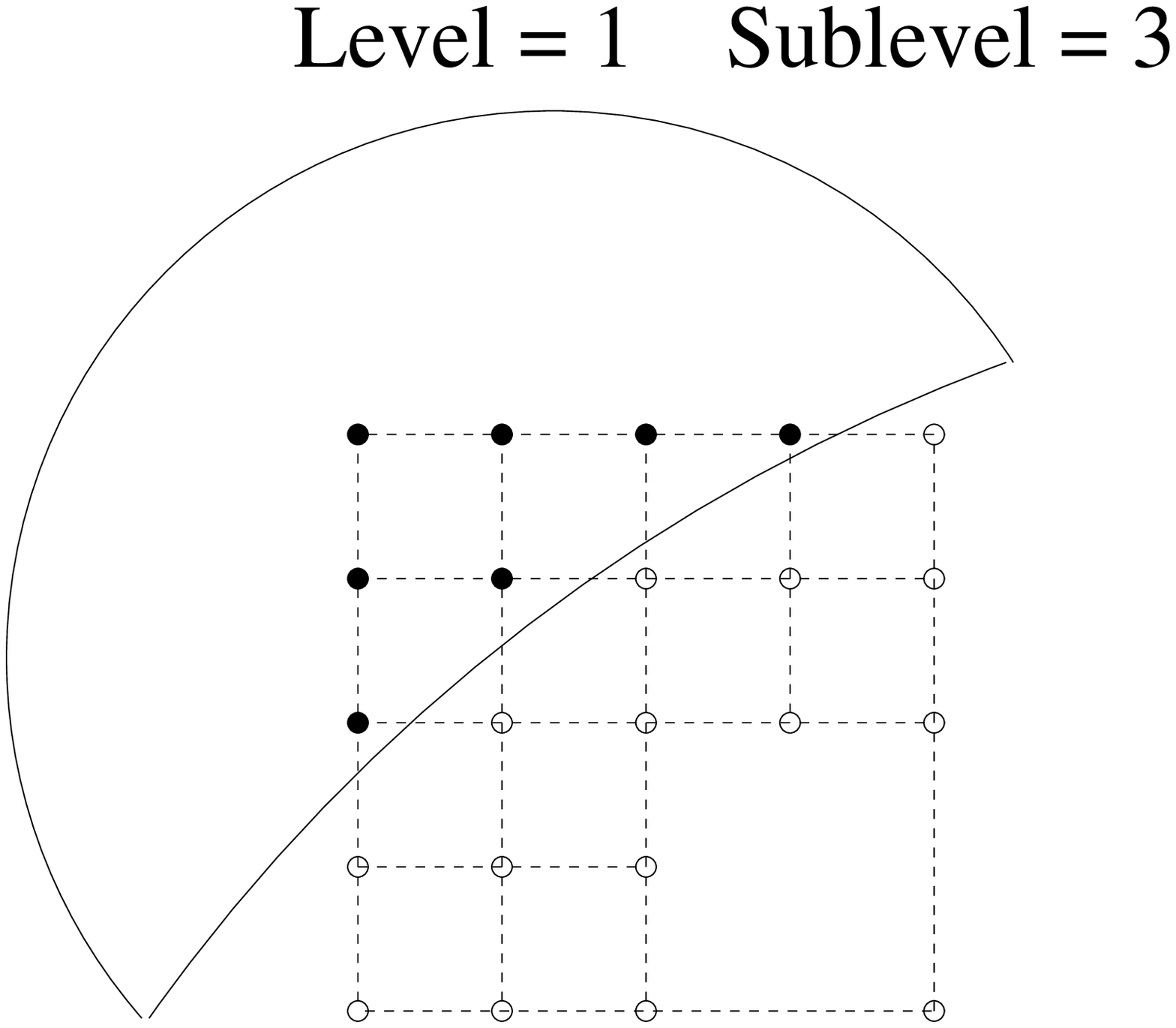}
  \includegraphics[width=0.45\textwidth]{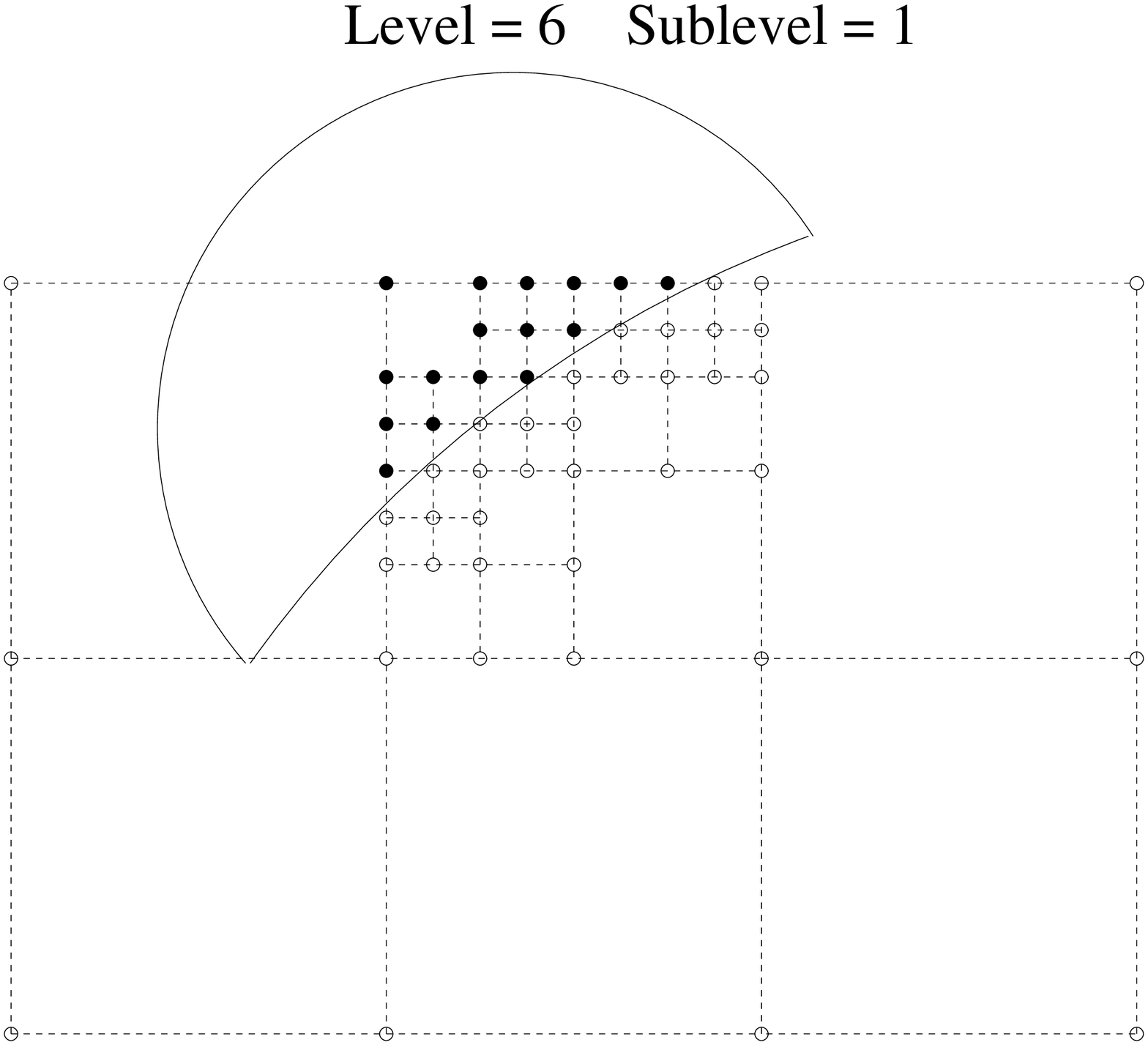}
  \includegraphics[width=0.45\textwidth]{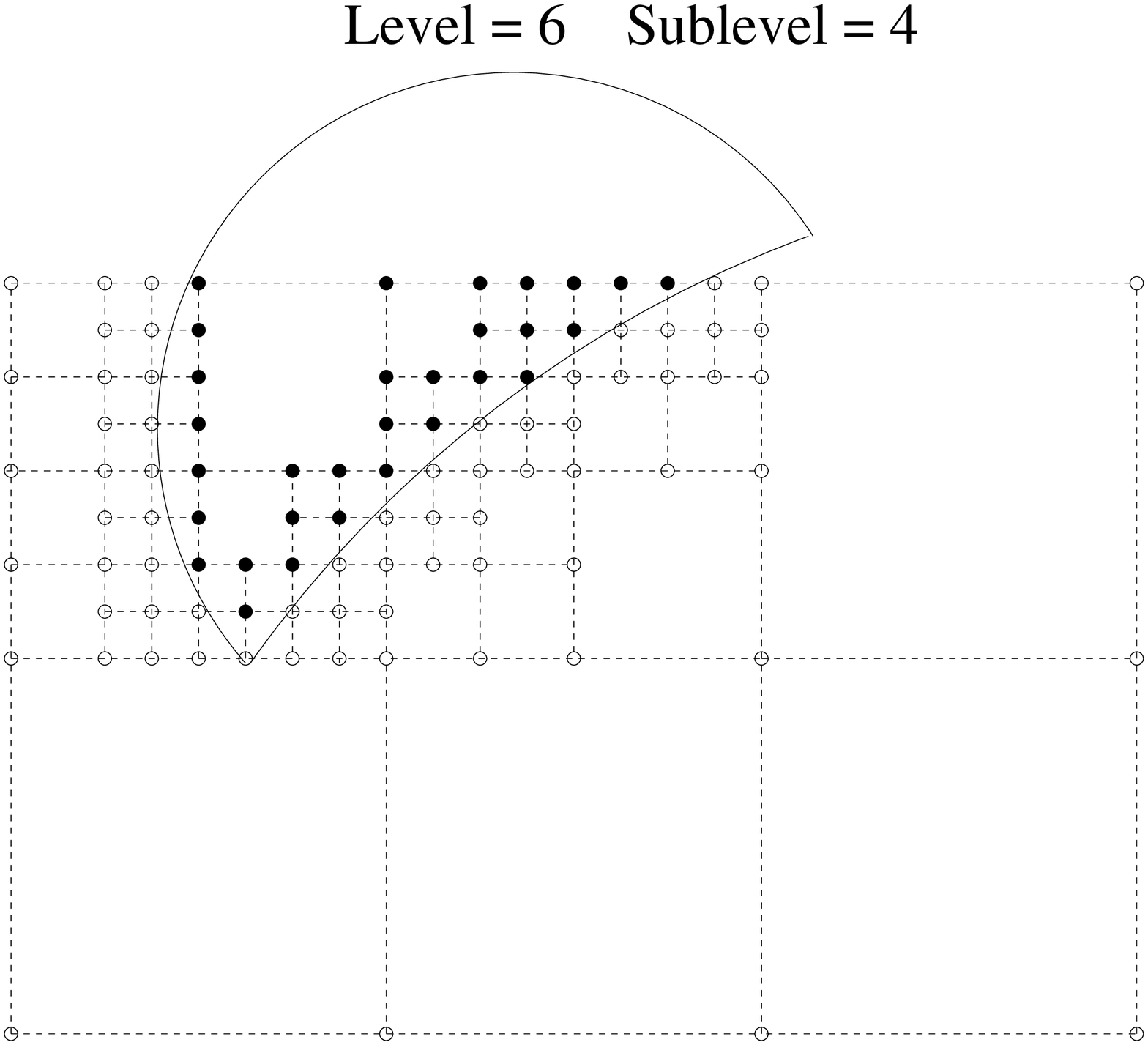}
  \caption{
A schematic representation of the tiling algorithm.  The algorithm uses a 
macrogrid with adaptive
resolution that subdivides if the number of corners that are ``in''
is $\neq$ 0 or 4.  Once the maximum number of sublevels is reached,
the macrogrid is shifted in a spiraling geometry about the origin.
}
  \label{fig:inout}
  \end{center}
\ep
A grid point $(x,y)$ is flagged as ``in'' if,
\be
  [x - \alpha_x(x,y)]^2 + [y - \alpha_y(x,y)]^2 \leq \Rsrc^2
  \label{eq:in}
\ee
and ``out'' if,
\be
  [x - \alpha_x(x,y)]^2 + [y - \alpha_y(x,y)]^2 > \Rsrc^2
  \label{eq:out}
\ee
Here $\vec{\alpha}(x,y)$ is the deflection angle due to the SIE at point 
$(x,y)$ and relates the image plane grid points to the source plane 
coordinates $(u,v)$ via the lens equation,
\begin{eqnarray}
  u & = & x - \alpha_x(x,y) \nonumber \\
  v & = & y - \alpha_y(x,y)
\end{eqnarray}
We then recursively subdivide portions of this grid until all four 
corners of each subgrid are either ``in'' or ``out'' (or until the 
maximum number of subdivisions has been reached).  For a given subgrid 
this implies that our criterion for further subdivision is:
\be
  \begin{array}{ccll}
  N_{in} & = & \mbox{ 1, 2, or 3} & \mbox{ : subdivide} \\
  N_{in} & = & \mbox{ 0 or 4}     & \mbox{ : do not subdivide}
  \end{array}
\ee
where $N_{in}$ is the number of corners that are flagged as being ``in'' 
the image.

Once this initial grid has been appropriately subdivided, we then 
move the macrogrid in a spiraling geometry and repeat the process (see 
Figure \ref{fig:spiral}) keeping a running total of the image area upon 
each revolution of the spiral.
\bp
  \begin{center}
  \includegraphics[width=0.6\textwidth]{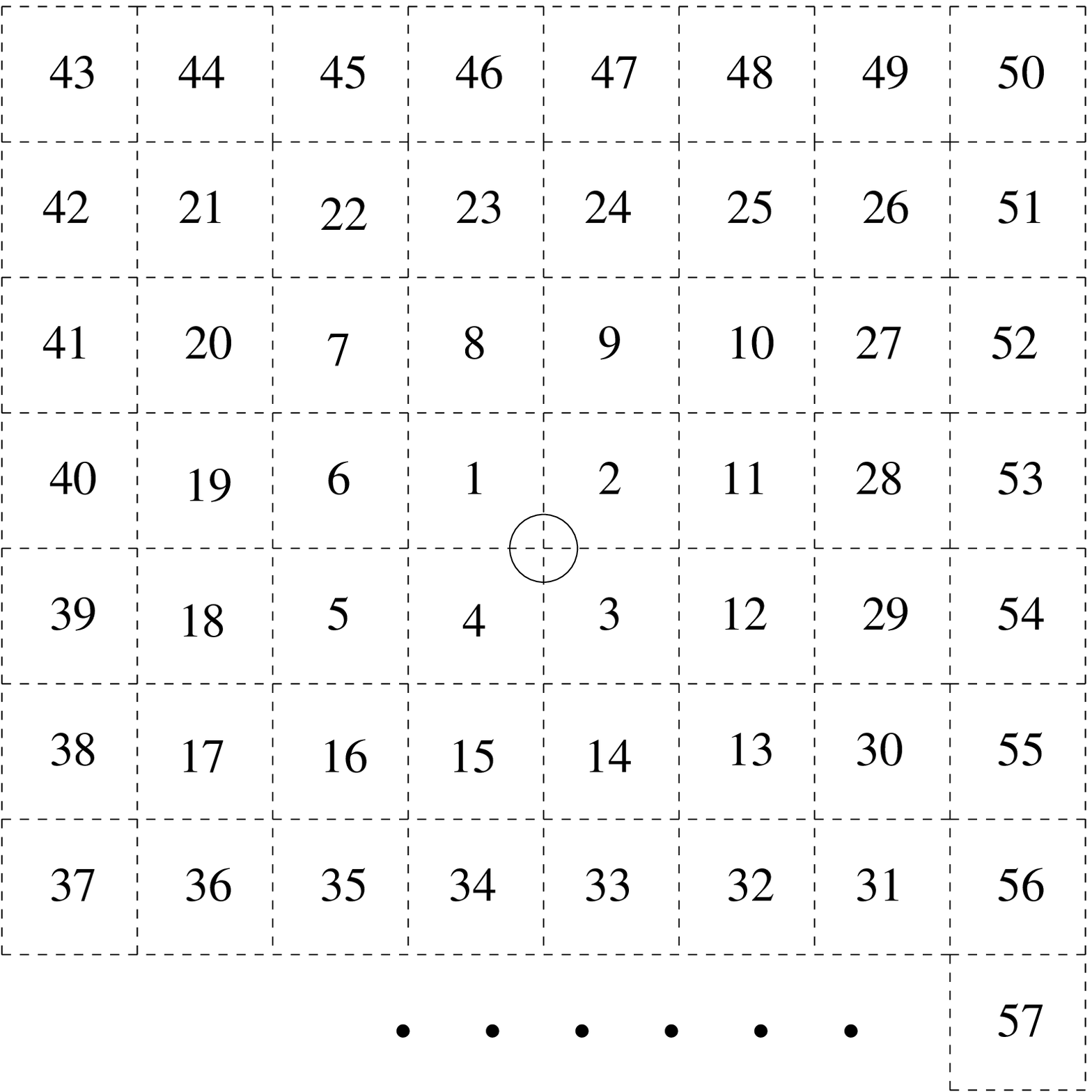}
  \caption{
A schematic representation of the evaluation order for the ``steps''
for the spiraling algorithm.  The origin is marked with an $O$ and the 
spiraling terminates once $N_{tol}$ revolutions yield no significant
change in the integrated area.
}
  \label{fig:spiral}
  \end{center}
\ep
This spiraling procedure self-terminates once the macrogrid has completed 
$N_{tol}$ revolutions with little to no change in the total image area.

Our technique has the advantage of achieving high resolution with 
relatively few subdivisions per grid and, more importantly, does not 
require preset boundary conditions.  That is, the self-termination 
criterion simply stops the spiraling procedure once the total image area 
is no longer changing.  However, care must be taken to avoid missing 
``distant'' images (images which appear far from the origin), and in 
practice we do set a minimum number of revolutions for the spiral.

In the previous example, we assumed our function evaluations at the grid points 
took only two values: 1 or 0 (i.e., ``in'' or ``out'' of the image).  The next 
step is to allow each grid point to assume a continuous value $f(x,y)$.  With 
this generalization, we must modify our subdivision criterion which we now take 
to be,
\be
  \begin{array}{cccl}
  |F_{avg} - F_c| & \geq & \epsilon \ |F_{avg}|& 
  \mbox{ : subdivide} \\
  |F_{avg} - F_c| & < & \epsilon \ |F_{avg}| & 
  \mbox{ : do not subdivide},
  \end{array}
\ee
where
\be
  F_{avg} = \frac{1}{4} [f(x_1,y_1)+f(x_2,y_1)+f(x_1,y_2)+f(x_2,y_2)]
\ee
is the average of the function values at the four grid corners and,
\be
  F_c = f\left(\frac{x_1+x_2}{2},\frac{y_1+y_2}{2}\right)
\ee
is the function value at the center of the grid.  The tolerance value 
$\epsilon$ must be set by hand and is roughly a measure of how much 
$f(x,y)$ varies over the region $[x_1:x_2][y_1:y_2]$.  That is, our grid 
only increases resolution in regions where the function varies rapidly 
with either $x$ or $y$.  Since the spiral termination criterion is still 
such that the total integral does not significantly change after $N_{tol}$ 
revolutions, this 2-D integration scheme is ideally suited to integrate 
functions for which $f(x,y) \rightarrow 0$ as $x,y \rightarrow \infty$.

\end{document}